# Astrobiological Potential of Venus Atmosphere Chemical Anomalies and Other Unexplained Cloud Properties

**Janusz J. Petkowski**[1,2,3]*, **Sara Seager**[1,4,5], **David H. Grinspoon**[6], **William Bains**[1,7], **Sukrit Ranjan**[8], **Paul B. Rimmer**[9,10,11], **Weston P. Buchanan**[1,12], **Rachana Agrawal**[1], **Rakesh Mogul**[13] and **Christopher E. Carr**[14]

[1] Department of Earth, Atmospheric and Planetary Sciences, Massachusetts Institute of Technology, 77 Massachusetts Avenue., Cambridge, MA 02139, USA
[2] Faculty of Environmental Engineering, Wroclaw University of Science and Technology, 50-370 Wroclaw, Poland
[3] JJ Scientific, Mazowieckie, 02-792 Warsaw, Poland
[4] Department of Physics, Massachusetts Institute of Technology, 77 Massachusetts Avenue., Cambridge, MA 02139, USA
[5] Department of Aeronautics and Astronautics, Massachusetts Institute of Technology, 77 Massachusetts Avenue., Cambridge, MA 02139, USA
[6] Planetary Science Institute, 1700 East Fort Lowell, Suite 106, Tucson, AZ 85719-2395, USA
[7] School of Physics & Astronomy, Cardiff University, 4 The Parade, Cardiff CF24 3AA, UK
[8] Lunar & Planetary Laboratory, Department of Planetary Sciences, University of Arizona, Tucson, AZ, USA
[9] Department of Earth Sciences, University of Cambridge, Downing Street, Cambridge CB2 3EQ, UK
[10] Cavendish Laboratory, University of Cambridge, JJ Thomson Avenue, Cambridge CB3 0HE, UK
[11] MRC Laboratory of Molecular Biology, Francis Crick Avenue, Cambridge CB2 0QH, UK
[12] School of Aeronautics and Astronautics, Purdue University, 701 W. Stadium Ave., West Lafayette, IN 47907, USA
[13] California Polytechnic University, Pomona, Pomona, CA, USA
[14] School of Aerospace Engineering and School of Earth and Atmospheric Sciences, Georgia Institute of Technology, Atlanta, GA, USA

* Correspondence: Janusz J. Petkowski, jjpetkow@mit.edu

**Abstract:** Long-standing unexplained Venus atmosphere observations and chemical anomalies point to unknown chemistry but also leave room for the possibility of life. The unexplained observations include several gases out of thermodynamic equilibrium (e.g. tens of ppm $O_2$, the possible presence of $PH_3$ and $NH_3$, $SO_2$ and $H_2O$ vertical abundance profiles), an unknown composition of large, lower cloud particles, and the "unknown absorber(s)". Here we first review relevant properties of the Venus atmosphere and then describe the atmospheric chemical anomalies and how they motivate future astrobiology missions to Venus.

**Keywords:** Venus clouds; atmospheric chemistry; acidity; habitability; atmospheric gases

## 1. Introduction

Scientists have been speculating on Venus as a habitable world for over half a century (e.g. (Bains et al., 2023b, 2021a; Cockell, 1999; Dartnell et al., 2015; Grinspoon and Bullock, 2007; Izenberg et al., 2021; Kotsyurbenko et al., 2021; Limaye et al., 2021b, 2018; Mogul et al., 2021a; Morowitz and Sagan, 1967; Patel et al., 2021; Schulze-Makuch et al., 2004; Seager et al., 2021)), based on the Earth-like temperature and pressure in Venus' clouds at 48–60 km above the surface. The hypothesis that Venusian clouds may be inhabited by an aerial biosphere has recently been bolstered by a tentative detection of the gas phosphine ($PH_3$) in the atmosphere of Venus (Bains et al., 2021c; Greaves et al., 2021b). Phosphine's presence at ppb levels is not explained by any known chemistry (Bains et al., 2021c, 2022c, 2022d, 2022a, 2023a). $PH_3$, however, is not the only Venus' atmospheric constituent that suggests unknown chemical processes in the clouds and leaves room for the possibility of life. The presence of such unexplained chemicals came to the forefront due to recent efforts to re-analyze and reinterpret the legacy data collected by both the Pioneer Venus and Venera probes (Bains et al., 2021a; Mogul et al., 2021b, 2021a).

The former Soviet Union has sent thirteen successful in situ Venus probe missions (between 1967 and 1984). Eleven of them (Venera 4–14) were atmospheric probes and landers, two were balloons, as well as atmospheric probes and landers (VeGa 1–2). The United States launched three flyby missions and a single large mission, Pioneer Venus, in 1978, with orbiter and four in situ atmospheric probes (Fimmel, 1983).

In this paper, we review and summarize Venus' long-lasting, unexplained atmospheric observations, which have been acquired over the span of the last half century. We focus on detections and observations that have been previously dismissed as artifactual, forgotten, or otherwise remained unexplained for decades. Such unexplained observations include, for example, the "unknown absorber(s)" and the chemical composition and shape of Mode 3 cloud particles. The tentative, dismissed, unexplored, or forgotten chemical atmospheric constituents include tens of ppm $O_2$, the possible presence of organics, $PH_3$ and $NH_3$, to name a few. We also discuss the anomalous vertical abundance profiles of $SO_2$ and $H_2O$ and summarize the model that could explain them, emphasizing the model's astrobiological implications. We discuss the original observations and methods used and the validity of the original discoveries. Such unexplained, unexplored, and chemically anomalous properties motivate and justify a dedicated Venus mission to confirm previous measurements with modern instrumentation and test for possible mechanisms behind the legacy observations, including the presence of life.

## 2. Motivation for a Venus Astrobiology Mission

There are many scientific reasons to explore the atmosphere, surface, and clouds of Venus, and many possible mission architectures and instrumented platforms to make measurements (Limaye and Garvin, 2023). The basic motivation to confirm and study unexplored, unexplained, and anomalous measurements in the Venusian atmosphere is that it is through detailed studies of such anomalies and their context that they can be validated and their explanation discovered, potentially including the presence of life (Cleland, 2019a, 2019b). This broad approach has been illustrated by the Galileo experiment, that is, the attempt to detect evidence of life on Earth solely from remote observations from Galileo during its flyby of Earth (Sagan et al., 1993). Sagan et al. (1993) concluded that the results of the Galileo flyby were consistent with the existence of life on Earth, based on the identification of atmospheric anomalies (e.g., the coexistence of significant $CH_4$ and $O_2$) that were challenging to explain with known abiotic mechanisms and understanding of planetary physical properties but were possible to explain with biotic mechanisms. Life has been postulated as a potential source or contributor to observed, yet poorly constrained, Venus cloud properties (e.g., the source of the strong UV absorption, mode 3 particles, etc.), yet Venus is much less well understood than Earth. Executing a similar procedure for Venus will require resolving current mysteries or unknowns regarding its atmosphere, identifying anomalies that persist despite improved understanding of the general atmospheric state, and seeking and testing explanations for these anomalies and other unexplained observations. Upon detailed study of possible missions focused on astrobiology and, in particular, on life detection, it has become clear that only a focused in situ mission and/or a sample return mission carries the likelihood of providing definitive answers to the crucial questions posed with regard to atmospheric chemistry, habitability of the clouds, and possible presence of life (Seager et al., 2022b).

This point is well illustrated by the fact that the two most recent Venus missions, both orbiters, have merely confirmed and deepened many of the outstanding mysteries of the Venus cloud region. Both Venus Express (ESA) (Svedhem et al., 2009) and Akatsuki (JAXA) (Nakamura et al., 2011) have been successful orbital spacecraft, which returned valuable data on the cloud composition and structure and on atmospheric dynamics and composition. These missions have continued to monitor the "unknown absorber(s)," which absorb a large fraction of the incident solar radiation, but have not succeeded in identifying the substance(s) responsible for this enormous unexplained absorption. Likewise, Venus Express and Akatsuki have generally confirmed the overall picture of the Venus clouds and cloud-level atmosphere provided by earlier American and Soviet entry probes (Pioneer Venus, the Venera and VeGa missions) and decades of ground-based observations. As a result, ESA and JAXA missions have filled in many details of cloud structure and dynamics, but they have not resolved the persistent mysteries that involve possible trace cloud components, unusual particle shapes, and trace atmospheric gases. These unexplained observations, both individually and taken together, are significant motivators to return to Venus for in situ observations.

The ill-defined Venus cloud properties and unexplored chemical observations fuel speculation about possible biological activity. Detailed characterization of cloud particle properties has proven particularly impervious to remote investigation and would require direct sampling of the clouds. Likewise, detection of biological activity or even life itself would require high-fidelity, novel in situ analytical methods or atmospheric sample return missions and cannot be accomplished using remote sensing techniques (Seager et al., 2022b). For an illustrative analogy, consider trying to make a definitive determination of the presence of life in a terrestrial location such as the Atacama Desert where microbial life is present but sparsely and at low abundance. Satellite remote sensing might hint at some of the right conditions, such as moisture and temperature range, but a definitive positive detection of life would likely require a platform that could directly sample the upper layers of the desert surface or even bring samples back to specialized laboratories for further study (see, e.g., Cabrol et al., 2007; Parro et al., 2011; Vítek et al., 2012). Such difficulties illustrate the limitations of remote sensing for biology by orbital missions. What is missing and the logical next step is direct sampling of the environment by an entry probe equipped with modern instrumentation (see, e.g., Limaye et al., 2021b; Schulze-Makuch and Irwin, 2002; Seager et al., 2022b, 2022a).

In this context, it is striking to consider that there has never been an in situ investigation of the atmosphere and clouds of Venus employing 21st century scientific instrumentation. The most recent American entry probes were the Pioneer Venus probes that flew in 1978. The Soviet VeGa balloons flew in 1985. It has been 38 years since any instrument from Earth was flown to directly investigate the atmosphere and the clouds of Venus. The entire scientific field of Astrobiology has matured in the interim. We now know questions to ask that we could not have formulated in the 1980s, but even more important is the progress in scientific instrumentation and miniaturization of electronics during these decades.

## 3. Venus' Unexplained Observations and Understudied Cloud Properties

Many intriguing in situ observations of Venus have never been fully explored (Figure 1). Nearly all of these observations could be the result of biological activity, though life may not be required to explain any of them.

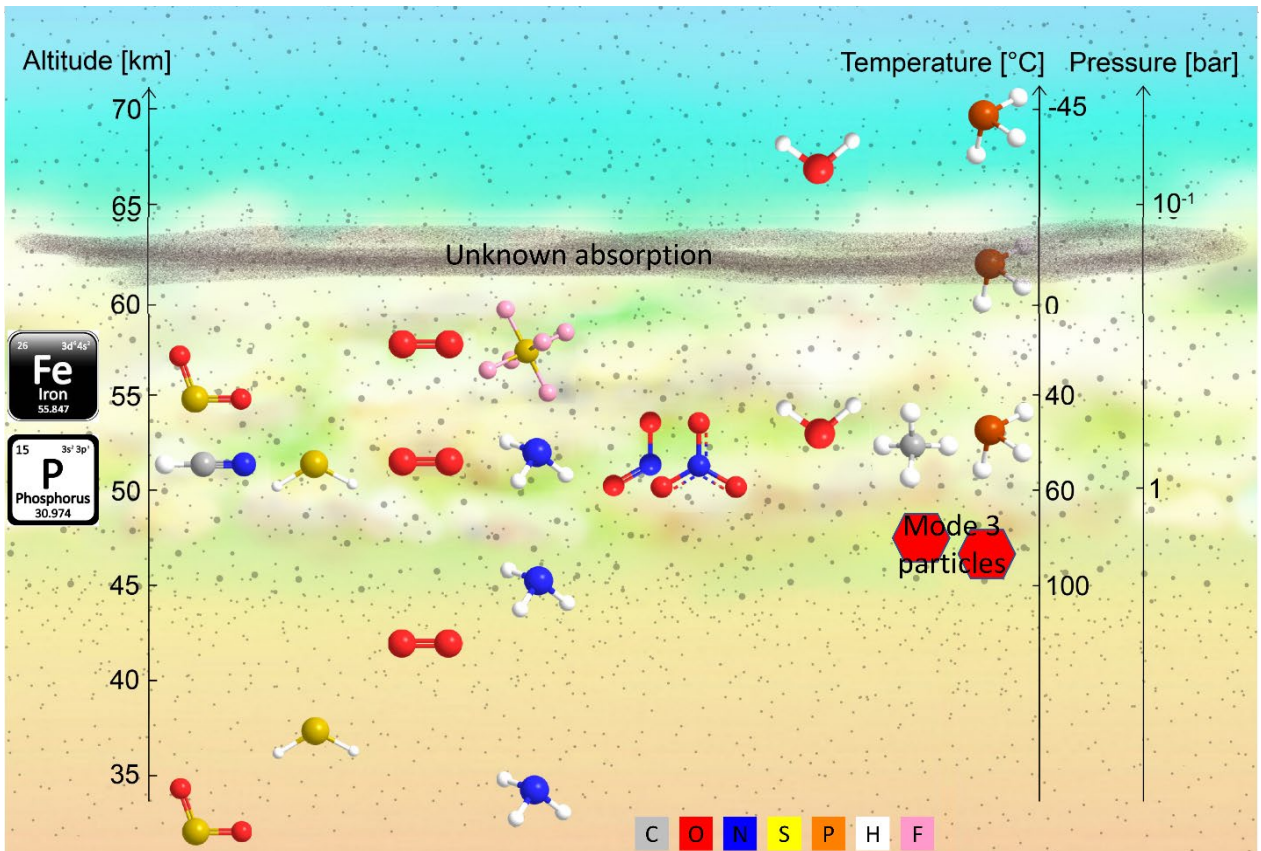

**Figure 1.** Unexplained and unexplored Venus' atmospheric observations. The molecule models show individual detections and altitudes for the atmospheric observations (e.g. $NH_3$ has been tentatively observed three times, twice, at two different altitudes by Venera 8, at ~32 and ~45 km, and once by Pioneer Venus at ~51 km, see also Table 3). If the observation has been made at the altitude range the molecule model is placed at the highest altitude of the range (e.g. $SF_6$ has been tentatively detected between 35 and 58 km). The $SO_2$ and $H_2O$ molecule models do not represent every individual observation but rather denote the anomalous abundance profiles for $SO_2$ and $H_2O$ in the atmosphere (see (Bains et al., 2021a; Rimmer et al., 2021) and Section 3.5). Most of the unexplained atmospheric observations have been recorded within the clouds (48-70 km) and in the stagnant haze layer below the clouds (31-47 km).

*3.1. The Unknown Absorber(s)*

While Venus appears relatively bland and featureless at visible wavelengths, observers starting in the 1920s noticed unusual high-contrast features in the ultraviolet (Ross, 1928). These features move with the ~4-day super-rotation of Venus' upper cloud deck, yet also display great variability on a wide range of temporal and spatial scales. Much effort has gone into attempting to identify the substance(s) responsible for the absorption between 320–400 nm, but no proposed candidate satisfies all of the observational constraints, leading to the oft-used descriptive term "unknown UV absorber." We note, however, that the absorption of radiation that is associated with the "unknown UV absorber" is not restricted exclusively to UV but also extends into longer wavelengths (Limaye et al., 2018; Pérez-Hoyos et al., 2018). Hence, throughout this article we use the term "unknown absorber" when referring to this unexplained phenomenon, knowing that multiple chemical species (*i.e.*, "unknown absorbers") could contribute to this phenomenon.

The current consensus, based, for example, on Venus photometric measurements, is that the unknown absorption occurs predominantly right below the cloud tops, at 60 km, and is associated with cloud particles (possibly with the smallest micron sized cloud particles referred to as Mode 1) rather than gaseous species (Ekonomov et al., 1984; Titov et al., 2012; Tomasko et al., 1980, 1979) (reviewed by Titov et al., 2018)).

After the upper clouds were identified as being composed primarily of sulfuric acid droplets (Hansen and Hovenier, 1974; Knollenberg and Hunten, 1980; Young, 1973), efforts to identify the absorber largely focused on sulfur compounds, including $SO_2$, $S_2O$, $S_2O_2$, and various allotropes of elemental sulfur ($S_3$, $S_4$ and $S_8$) (Table 1). Other proposals have focused on elemental chlorine ($Cl_2$), which has been identified in the upper atmosphere and shows absorption features in roughly the right spectral range. A summary of proposed candidates is given in the works of Limaye et al (2021a), Mills et al. (20070, and Pérez-Hoyos et al. (2018) and in Table 1 below.

Despite decades of effort and observations by two orbiting spacecraft in the 21st century (Venus Express by ESA and Akatsuki by JAXA), none of the proposed candidate molecules have been found to entirely fit the observational data. The candidate molecules either have too low abundance ($S_2O_2$) (Krasnopolsky, 2018; Titov et al., 2018) or do not entirely fit the spectral absorption profile ($FeCl_3$) (Pérez-Hoyos et al., 2018; Rustad and Gregory, 1977) (Table 1). In principle, one can overcome the problem that the narrow spectral absorption of some candidate molecules does not match the broad absorption of the unknown absorber by postulating that the absorber is a mix of materials and not a single chemical species. However, most of the proposed UV-absorbing species are unstable to UV photochemistry or predicted to be present at extremely low abundances in the absence of biological activity. Therefore, the mystery of the Venusian absorber persists. The unknown absorber is remarkably efficient, capturing more than 50% of the solar energy reaching Venus, with consequent effects on atmospheric structure and dynamics.

Several researchers have suggested that qualities of the unknown absorber could be a signature of biological activity in the clouds (Limaye et al., 2018; Schulze-Makuch et al., 2004). The spectral characteristics of the Venus clouds, including the strong UV absorption, are consistent with the spectrum of certain types of terrestrial bacteria (Limaye et al., 2018). The spatial and temporal patterns of the unknown absorber are somewhat reminiscent of terrestrial algal blooms (Grinspoon and Bullock, 2007; Limaye et al., 2018). The great efficiency of absorption, if utilized as a photosynthetic pigment, could provide a large amount of metabolic energy (Grinspoon, 1997). The consistency between the UV absorption spectrum of the unknown Venus absorber and those of bacterial cells is perhaps not surprising as many pigments and various aromatic compounds and proteins (especially in combination) present in bacterial cells have broad and diverse UV absorption properties. A selection of the spectra of UV-absorbing biomolecules would reproduce the absorber spectrum, but in the absence of any evidence that any of the components are present this would be an arbitrary fit and not evidence for the presence of those compounds.

Attempts have been made to tie the possible abundance of the unknown absorber to the expected biomass of the hypothetical aerial biosphere in Venus' clouds (Jordan et al., 2022). However, as also noted by Jordan et al. (2022), there is no reliable way to estimate and correlate the biomass abundance to the possible abundance of the unknown absorber, in a strict gram per gram fashion. This is because some biological pigments have extraordinarily strong absorption, much stronger than, for example, simple salts. Therefore, a strongly absorbing species is not inconsistent with a very low abundance biomass.

Recently, Benner and Spacek (2021), Spacek (2021), Spacek et al. (2023), and Spacek and Benner (2021) speculated on organic molecules inside the Venus cloud particles as the unknown absorber. The proposal comes from laboratory experiments that started with simple organic molecules, including formaldehyde, dissolved in concentrated sulfuric acid. A chain of chemical reactions led to a rich variety of yellow-, red-, and brown-colored organic molecules. The hypothesis is that the simple organic molecules that are precursors to the organics responsible for the unknown absorber originate from meteoritic delivery, photochemistry, or even possibly life itself (Benner and Spacek, 2021; Spacek, 2021; Spacek et al., 2023; Spacek and Benner, 2021).

Indeed, the petrochemical industry uses concentrated sulfuric acid as a catalyst during octane production from isobutane and butene and finds a rich chemistry in concentrated sulfuric acid from the reactivity of hydrocarbon molecules (Albright et al., 1972; Huang et al., 2015; Miron and Lee, 1963). While the resulting compounds called "red oil" are an undesirable side product, this chemistry substantiates the idea that the Venus cloud sulfuric acid particles can support diverse organic chemistry independent of the presence of life.

The direct detection of organic chemicals has never been attempted and should be a priority for future in situ missions to Venus. Neither Venera/VeGa probes nor Pioneer Venus directly searched for organic chemistry in the clouds. We will get more information on the possible composition of the unknown absorber(s) via NASA's DAVINCI orbiter's Compact Ultraviolet to Visible Imaging Spectrometer (CUVIS) instrument (Garvin et al., 2022), but a direct detection and identification of organic chemicals in the clouds of Venus is not a target of the selected NASA and ESA missions. The detection of organic chemicals within the cloud particles is, however, one of the science objectives of the Rocket Lab mission to Venus (French et al., 2022) and its science instrument, the autofluorescence nephelometer (AFN) (Baumgardner et al., 2022).

**Table 1.** List of candidates for the "unknown absorber(s)".

| Proposed candidate absorber molecule | Proposed absorber explanation | Cons |
|---|---|---|
| sulfur dioxide $SO_2$ | $SO_2$ gas is the main absorber at wavelengths from 200 to 320 nm (Ekonomov et al., 1983; Pollack et al., 1980, 1979). | The absorption at wavelengths longer than 320 nm cannot be accounted for by $SO_2$ (Blackie et al., 2011; Pérez-Hoyos et al., 2018). |
| elemental sulfur allotropes $S_3$, $S_4$, $S_8$ | Various forms of gaseous and aerosol forms of elemental sulfur have been postulated to contribute to the absorber spectra as well as to the pale yellow color of Venus: $S_3$ and $S_4$ (Toon et al., 1982) $S_8$ (Hapke and Nelson, 1975; Schulze-Makuch and Irwin, 2006). | Sulfur aerosols ($S_8$) alone cannot account for the unknown absorber, as their abundance is too low at the cloud tops and its absorption profile does not agree with the absorber profile (Krasnopolsky, 2016, 2013); $S_3$ and $S_4$ absorption is centered at longer wavelengths, >360 nm (Pérez-Hoyos et al., 2018). |
| disulfur monoxide $S_2O$ | An irradiated version of $S_2O$ could contribute to the 350 nm core absorption feature and some absorption in the range of 400–500 nm (Hapke and Graham, 1989). Both gaseous and condensed phase $S_2O$ have been proposed to contribute to the unknown absorber spectra. | Spectral characteristics of *cyc*-$S_2O$ do not adequately match the unknown absorber spectra (Frandsen et al., 2020). Other $S_2O$ isomers are too unstable (Frandsen et al., 2020). |
| disulfur dioxide $S_2O_2$ | The *cis*- and *trans*- isomers of $S_2O_2$ (OSSO) are a good fit to the unknown absorber spectra (Frandsen et al., 2020, 2016; Wu et al., 2018). | Very short photochemical lifetime (seconds) of the OSSO species precludes its existence on the day side of the planet (Titov et al., 2018). Photochemical modeling of the atmosphere does not support $S_2O_2$ as major absorber (Krasnopolsky, 2018). |
| ammonia pyrosulfite $(NH_4)_2S_2O_5$ | Ammonia pyrosulfite aerosols may contribute to the near UV absorption if $(NH_4)_2S_2O_5$ forms in the cold top cloud regions (Titov, 1983). | Inconsistent with Pioneer Venus spectroscopic observations at 365 nm (Krasnopolsky, 1986). Venus clouds should have been brighter in UV at low altitudes (Krasnopolsky, 1985). |
| sulfur dichloride $SCl_2$ | Proposed to account for the core absorption around 350 nm (Krasnopolsky, 1986). | Too narrow absorption to account for the entire unknown absorber (Pérez-Hoyos et al., 2018). Photochemical abundance estimates are too low to account for the unknown absorption (Krasnopolsky, 1986). Other postulated chlorine-sulfur species (e.g. $SO_2Cl$, $SOCl_2$, $SO_2Cl_2$, $SO_4Cl$) (Baines and Delitsky, 2013) have not been observed or have very short photochemical lifetime (Krasnopolsky, 2013, 2007). |
| perchloric acid $HClO_4$ | Suggested as a component of aerosols contributing to the unknown absorber (Von Zahn et al., 1983). Recent re-analysis of the Pioneer Venus LNMS data shows evidence of oxychlorine species, e.g. chlorous acid ($HClO_2$), in the atmosphere, their abundance however is uncertain (Mogul et al., 2021b). | Not good fit to the Venus spectra. The production rate of $HClO_4$ should be negligible, not enough to account for the observed unknown absorber (Krasnopolsky, 1986). |
| hydrobromic acid $HBr$ | Proposed to account for the core absorption around 350 nm (Sill, 1975). | Not good fit to the Venus spectra. No confirmed detection; upper limits of ~1 ppb at the cloud tops give abundances too low for $HBr$ to be a major contributor to the unknown absorber (Krasnopolsky and Belyaev, 2017). |
| chlorine $Cl_2$ | Proposed to account for the core absorption around 350 nm (Pollack et al., 1980). | Too narrow absorption to account for the entire unknown absorber (Pérez-Hoyos et al., 2018). Photochemical abundance estimates are too low to account for the unknown absorption (Krasnopolsky, 1986). |
| iron chloride FeCl$_3$ in aerosols | Proposed to account for the core absorption around 350 nm (Krasnopolsky, 2017, 1985; Zasova et al., 1981). | Too narrow absorption to account for the entire unknown absorber (Pérez-Hoyos et al., 2018; Rustad and Gregory, 1977). |
| nitrosylsulfuric acid NOHSO$_4$ | Together with other species proposed to qualitatively explain Venus' UV albedo (Krasnopolsky, 1986; Watson et al., 1979). | Not good fit to the Venus spectra. The predicted abundance of NOHSO$_4$ is insufficient to account for the observed unknown absorber (Krasnopolsky, 1986). |
| nitric oxide NO and other NO$_x$ species | Postulated to contribute to the unknown absorber (Shaya and Caldwell, 1976). | Cannot fully explain the unknown absorber. Photochemical abundance estimates are too low to account for the unknown absorption (Krasnopolsky, 1986). See also early ground-based observations upper limits on NO$_x$ abundance above the clouds (see (Moroz, 1981), their Table VI) and the newest upper limits on NO$_x$ in the Venus lower-mesosphere using SOIR on board Venus Express (Mahieux et al., 2023). |
| carbon disulfide CS$_2$ | Absorbs strongest at wavelengths <360 nm, but not between 330 and 600 nm (Keller-Rudek et al., 2013). | UV absorption at wavelengths longer than 330 nm cannot be accounted for by CS$_2$. |
| carbonyl sulfide OCS | Absorbs strongest at wavelengths <300 nm, but not between 330 and 600 nm (Keller-Rudek et al., 2013). | UV absorption at wavelengths longer than 330 nm cannot be accounted for by OCS. |
| fine graphite grains | Carbon suboxide polymer (which has a yellow color) and fine graphite grains have absorption bands in the UV (Shimizu, 1977). | Not good fit to the Venus spectra. |
| croconic acid | Sulfuric acid aerosols mixed with croconic acid; (Hartley et al., 1989) have been first to propose organic molecules as an unknown absorber candidate. | Not good fit to the Venus spectra (Bertaux et al., 1996). No evidence of croconic acid in Venus atmosphere (Mills et al., 2007). |
| complex organic chemicals | A mixture of diverse colored species of complex organic chemicals dissolved in sulfuric acid provides good fit to the absorber spectra and explains the pale yellow color of Venus (Spacek, 2021). | No direct evidence for sufficient amounts of organic chemicals in the top clouds, although the possibility that the absorber is brought to the upper clouds from lower atmospheric layers remains (Titov et al., 2018). |
| biomolecules and life itself | Some mixtures of UV absorbing biomolecules could reproduce the unknown absorber spectrum (Limaye et al., 2018). | Requires biological activity to explain the observed features of the unknown absorber. |

*3.2. Mode 3 Particle Composition*

The composition of a subset of Venus cloud particles, large particles (>7 μm in diameter) in the lower clouds called "Mode 3," is unknown (see Table 2 for a summary of Venus cloud particle properties vs. altitude, a summary discussion on the Mode 3 particles in (Bains et al., 2021a) and (Mills et al., 2007)). Adding to the mystery is the fact that the Mode 3 particles as measured by the Large Cloud Particle Size Spectrometer (LCPS) onboard the Pioneer Venus Large Probe appear to be non-spherical (Knollenberg and Hunten, 1980, 1979). Data from the Pioneer Venus Optical Array Spectrometer (OAS) (Esposito et al., 1983) also support non-spherical particles. The OAS instrument had three photodiode arrays that measured the shadows of passing particles, which makes the particle size measurement independent of particle composition. "Non-spherical" means the Mode 3 particles cannot be liquid droplets.

The nature and composition of the Mode 3 particles is debated with data presently in hand. The key derived parameter is refractive index, which comes from the Pioneer Venus nephelometer, which measured backscattered light in a range of angles. The refractive index of the particles in the lower clouds at 49 km is reported at 1.32 ± 0.03 assuming spherical droplets (Knollenberg et al., 1980; Ragent and Blamont, 1979). This value is lower than any plausible value for sulfuric acid, and therefore, it implies that the cloud particles located at these altitudes are not composed of pure concentrated liquid sulfuric acid. This result could indicate unknown chemistry and is intriguing with regard to the possible presence of "life as we know it," which cannot withstand a concentrated sulfuric acid environment. A possible explanation is non-spherical particles (Knollenberg and Hunten, 1980; Ragent and Blamont, 1979), which again imply non-liquid particles.

Several studies have questioned the existence of the large Mode 3 particles altogether and claimed, for example, that Mode 3 could be a large "tail" of the liquid Mode 2 particle distribution, once calibration errors were taken into account (James et al., 1997; Toon et al., 1984; Zasova et al., 1996). The possibility also remains that the OAS instrument could have, on occasion, measured overlapping shadows of two or more particles as they passed in front of the photodiode arrays. The subsequent re-examination of the evidence for the large solid Mode 3 particles in the clouds reaffirms the existence of the third large mode of particles (Knollenberg, 1984). However, their putative non-spherical, crystalline nature remains uncertain and can only be resolved with new in situ measurements (Knollenberg, 1984). See also an excellent summary of the Mode 3 particle debate in the work of Mills et al. (2007).

The unknown composition of the Mode 3 particles leaves room, albeit speculative, for unknown chemistry or life. Microbial cells within the droplets would cause an index of refraction discrepancy (analogously to bacteria in water (Waltham et al., 1994)). Alternatively, salt formation in a droplet, as a result of acid neutralization either through biological activity (Bains et al., 2021a) (see Section 4) or through incorporation of mineral dust lofted from the surface (Rimmer et al., 2021) would alter droplet composition away from pure concentrated $H_2SO_4$ to a more clement chemical environment with a different refractive index (Bains et al., 2021a; Mogul et al., 2021a).

Such decades-long lingering questions on the true nature of the Venus cloud particles should motivate new missions to focus on characterizing the Mode 3 cloud particles and the composition of the clouds and cloud aerosols in general.

**Table 2.** Characteristics of the Venusian cloud particles. Data from (Knollenberg et al., 1980; Knollenberg, 1982). (1) – Mode 1 particles; (2) – Mode 2 particles; (3) – Mode 3 particles. Mode 1 particles have mean diameter around 0.4 µm, Mode 2 have diameter of few µm, Mode 3 particles are larger than Mode 1 and Mode 2 and have diameters >7 µm.

| Region | Altitude (km) | Temperature (K) | Pressure (atm) | Cloud Particle Properties | | |
|---|---|---|---|---|---|---|
| | | | | Average Num. Density (n cm$^{-3}$) | Mean Diameter (µm) | Consensus Particle Composition |
| Layers above upper haze | 100–110 | | | | | N/A |
| Upper haze | 70–90 | 225–190 | 0.04–0.0004 | 500 | 0.4; Bimodal (Venus Express) | 70% $H_2SO_4$ 30% $H_2O$ (if present); Unknown |
| Upper cloud | 56.5–70 | 286–225 | 0.5–0.04 | (1)–1500 (2)–50 | Bimodal 0.4 and 2.0 | liquid 80% $H_2SO_4$ 20% $H_2O$ |
| Middle cloud | 50.5–56.5 | 345–286 | 1.0–0.5 | (1)–300 (2)–50 (3)–10 | Trimodal 0.3, 2.5 and 7.0 | liquid 90% $H_2SO_4$ 10% $H_2O$ |
| Lower cloud | 47.5–50.5 | 367–345 | 1.5–1.0 | (1)–1200 (2)–50 (3)–50 | Trimodal 0.4, 2.0 and 8.0 | liquid 98% $H_2SO_4$ 2% $H_2O$ (or fuming acid; $H_2SO_4$ + $SO_3$)) |
| Lower haze | 31–47.5 | 482–367 | 9.5–1.5 | 2–20 | 0.2 | Unknown |
| Sub-cloud layers | 46 and 47.5 | 378 and 367 | 1.8–1.5 | 50 and 150 | Bimodal 0.3 and 2.0 | Unknown |

*3.3. Presence of Non-Volatile Elements in the Cloud Particles*

Both the VeGa balloons' and Venera probes' in situ measurements of the elemental composition of the cloud particles suggest that non-volatile elements relevant for habitability are present. Venera 13 and Venera 14 analysis of cloud particles indicates the presence of sulfur, chlorine, and iron (Petrianov et al., 1981). VeGa 1 and 2 measurements of the cloud material by X-ray fluorescence spectrometer (XRF) suggest the significant presence of chlorine, sulfur (Surkov et al., 1986), and phosphorus (P) in the lower cloud (Andreichikov, 1987b), but little iron (in contrast to the Venera probe measurements). Other elements suspected to exist are I, Br, Al, Se, Te, Hg, Pb, Al, Sb, and As (Marov and Grinspoon, 1998). Indeed, recent preliminary re-analysis of Pioneer Venus LNMS data shows evidence of a non-homogenous composition of cloud and haze particles (Zolotov et al., 2023). The particles could contain many chemicals dissolved in concentrated sulfuric acid, for example, various metal ions, salts, silica and even "insoluble organics" (Zolotov et al., 2023).

Life as we know it requires metals, for example transition metals such as iron (Fe), and other non-volatile species for catalysis. Even for some of the most ancient enzymes, the protein's primary role appears to be to hold catalytic metals in place to facilitate a reaction. Detection of metals and other non-volatile species as components of cloud particles would support the potential for habitability of the Venus clouds. In other words, the presence of metals and other non-volatile elements is not a biosignature but is an indicator of habitability.

In the altitude range of 52 to 47 km, the abundance of phosphorus appears to be on the same order as the abundance of sulfur (Andreichikov, 1987a, 1987b). Phosphorus is most plausibly in the form of P(V) acids or oxides, such as $H_3PO_4$, $H_4P_2O_7$, etc. (Bains et al., 2021c; Krasnopolsky, 1989). If the Venera descent probe data are correct and some cloud particles indeed contain > 50% phosphorus species by mass, then by definition the concentration of sulfuric acid in those droplets must be < 50% (see Section 4 for further discussion of the composition of cloud particles). Above 52 km, no phosphorus was detected. It is, therefore, plausible that phosphorus is present in a condensed liquid or solid phase predominantly in the lower cloud layer (Bains et al., 2021c).

In summary, numerous early measurements by the VeGa balloons and the Venera probes suggest the cloud particles are not pure sulfuric acid and the particles likely contain a plethora of other dissolved species (e.g., molecules containing Fe, Cl, P, and others). The exact composition and the concentration of the dissolved species is unknown.

Establishing the presence of non-volatile elements or compounds in the cloud particles should be one of the main science objectives of any Venus mission focusing on cloud habitability and composition. The minimal objective for such missions should be to establish the elemental composition of cloud aerosols to ppb abundance levels, focusing on confirmation of the early measurements of Fe and P by the Venera and VeGa probes, and lighter metals and non-metal elements C, N, and O, including Si, which would have been a great tracer of silica-containing dust[1]. The identification of the parent compounds of the detected non-volatile elements should follow. In particular, the search for a large fraction of liquid or solid phase of phosphoric acid(s) or phosphate salts in the lower clouds is paramount. Such analysis would also address a range of non-biological issues with understanding the trace chemistry of the atmosphere, such as whether phosphorus species are really present (Krasnopolsky, 1989) and the possible presence of $FeCl_3$ and other metal-containing chemicals (Krasnopolsky, 2017).

*3.4. Unexpected Atmospheric Gases and Gas Vertical Abundances*

A number of trace gases with unexplained abundance profiles have been observed to exist in the atmosphere of Venus (Table 3). Some of them (*e.g.*, $O_2$ or $NH_3$), aside from being relevant as potential signs of life in their own right, indicate chemical disequilibrium when considered together with the main atmosphere constituents. Earth's atmospheric disequilibrium is a result of life's activity, as exemplified by the coexistence of $N_2$ and $O_2$ (Krissansen-Totton et al., 2016). Although Venus' atmosphere is not as far from equilibrium as Earth's atmosphere is, the trace gas species detected at Venus indicate chemical disequilibrium in the clouds. Those gases include CO, $SO_2$, $H_2O$, $S_x$, OCS, and $H_2$ (Von Zahn and Moroz, 1985), as well as additional trace gas species detected in situ by the Venera and Pioneer probes (including those identified in the recent reanalysis of the Pioneer Venus LNMS data) such as $O_2$, $HNO_2$, $PH_3$, $H_2S$, $NH_3$, HCN (Table 3).

---

[1] VeGa X-ray fluorescence system could not measure elements lighter than phosphorus, Z <15, (it was designed to do so, but instrumental issues affected the reliability of the measurement). Therefore, for example, even if there were high loadings of organics, ammonium salts or silica-containing dust in the cloud particles, VeGa would not have detected them. See also (Krasnopolsky, 1989) for the detailed discussion of the VeGa mission results.

We note that a number of anomalous, unexplained findings of gases or gas distributions in Venus' atmosphere have been discounted because no explanation for their presence could be found (*e.g.*, for $O_2$ (Von Zahn et al., 1983)). We consider this argument weak; the measurements should be critically evaluated on instrumental and repeatability grounds, and explanations for robust measurements should then follow, not the other way around. For this reason, we next discuss some trace gas detections in detail.

**Table 3.** Measured abundances of trace gas species of interest in the Venus clouds and below-the-cloud atmosphere layers. Taken together, the gases demonstrate chemical disequilibrium in the Venus atmosphere. LNMS is the Pioneer Venus Large Probe Neutral Gas Mass Spectrometer. GC is the gas chromatograph on either Pioneer Venus or the Venera Probes. JCMT is James Clerk Maxwell Telescope, ALMA is Atacama Large Millimeter/submillimeter Array, SOFIA is Stratospheric Observatory for Infrared Astronomy.

| Gas | Observation | Altitude | Amount | Comments | Ref. |
|---|---|---|---|---|---|
| $PH_3$ | JCMT | >60 km | ~7 ppb | Tentative detection with Earth-based telescopes. | (Greaves et al., 2021b, 2021c, 2021a) |
| | ALMA | >60 km | ~7 ppb | Tentative detection with Earth-based telescopes. | (Greaves et al., 2021b, 2021a, 2021c) |
| | SOFIA | ~75 km | ~1 ppb | Tentative detection with Earth-based SOFIA telescope (Greaves et al., 2022a), compare with work by (Cordiner et al., 2022). | |
| | Pioneer Venus | 51 km | ~2 ppm | Identification in the re-analyzed Pioneer Venus LNMS data. | (Mogul et al., 2021b) |
| $NH_3$ | Venera 8 | 45 km | 0.01 % | Tentative detection by Venera 8 chemical probe at ~2 bar altitude. | (Surkov et al., 1973) |
| | Venera 8 | 32 km | 0.1 % | Tentative detection by Venera 8 chemical probe at ~8 bar altitude. | (Surkov et al., 1973) |
| | Pioneer Venus | 51 km | N/A | Tentative detection in the re-analyzed Pioneer Venus LNMS data. | (Mogul et al., 2021b) |
| $O_2$ | Venera 14 | 35–58 km | 18 ± 4 ppm | Detection by Venera 14 GC. | (Mukhin et al., 1982) |
| | Pioneer Venus | 52 km | 44 ± 25 ppm | Detection by Pioneer Venus GC. | (Oyama et al., 1980b) |
| | Pioneer Venus | 42 km | 16 ± 7 ppm | Detection by Pioneer Venus GC. | (Oyama et al., 1980b) |
| $H_2S$ | Venera 14 | 29–37 km | 80 ± 40 ppm | Detection by Venera 14 GC. | (Mukhin et al., 1982) |
| | Pioneer Venus | 51 km | N/A | Identification in the original, as well as re-analyzed Pioneer LNMS data. | (Mogul et al., 2021b) |
| | Pioneer Venus | <24 km | 3 ± 2 ppm | Identification in the original Pioneer LNMS data. | (Hoffman et al., 1980a) |
| HCN | Pioneer Venus | 51 km | N/A | Identification in the re-analyzed Pioneer LNMS data. | (Mogul et al., 2021b) |
| HNO$_2$ | Pioneer Venus | 51 km | N/A | Identification in the re-analyzed Pioneer LNMS data. | (Mogul et al., 2021b) |
| HNO$_3$ | Pioneer Venus | 51 km | N/A | Identification in the re-analyzed Pioneer LNMS data. | (Mogul et al., 2021b) |
| CH$_4$ | Pioneer Venus | 51 km | ~1000 ppm | Identified in the original and the re-analyzed Pioneer LNMS data; Possible contaminant. | (Donahue and Hodges Jr, 1993; Mogul et al., 2021b) |
| C$_2$H$_4$, C$_2$H$_6$, C$_6$H$_6$ | Pioneer Venus | 51 km | N/A | Identified in the re-analyzed Pioneer LNMS data; Possible contaminant. | (Mogul et al., 2021b) |
| SF$_6$ | Venera 14 | 35–58 km | 0.2 ± 0.1 ppm | Tentative detection by Venera 14 GC. | (Mukhin et al., 1982) |

**Oxygen (O$_2$).** In situ detections of O$_2$ in the Venusian lower clouds and below the clouds have been reported by at least two probes at the 10s of ppm level: Pioneer Venus (Oyama et al., 1980b) and Venera 13/14 (Mukhin et al., 1982) (Table 3). The Pioneer Venus Gas Chromatography (PVGC) (Oyama et al., 1980c) reported 43.6 ppm molecular oxygen (O$_2$) in the clouds at 51.6 km, 16 ppm below the clouds at 41.7 km, and no detection of O$_2$ at 21.6 km (Oyama et al., 1980b). Note that the PVGC preliminary gas measurements and O$_2$ abundance estimations of approximately 70 ppm (Oyama et al., 1979b) were revised on several occasions (Oyama et al., 1980a, 1979a) before the final PVGC gas abundances were published (Oyama et al., 1980b). The Venera 14 Gas Chromatograph (VGC) detected 18 ppm O$_2$ average between 35 and 58 km (Mukhin et al., 1982). The Venera 14 VGC O$_2$ abundance agrees with previously established Venera 12 VGC upper limits (<20 ppm) for O$_2$ below 42 km (Gelman et al., 1979a) and the final revised abundance recorded by the PVGC (16 ppm at 41.7 km) (Oyama et al., 1980b)[2].

The Neutral Mass Spectrometer (LNMS) on Pioneer Venus showed a signal of 32 amu, but this signal has been attributed to O$_2$ ions formed in the mass spectrometer from the reaction of CO$_2$ (Hoffman et al., 1980a) and, therefore, is considered unreliable as an indicator of the presence of O$_2$ in the atmosphere. The Venera 11 and Venera 12 mass spectrometers (VMS) detected an excess signal of 32 amu at altitudes of below 23 km (Istomin et al., 1980, 1979a, 1979b). As with LNMS, the reliability of this VMS measurement is uncertain. Venera 13 and Venera 14 mass spectra also show a mass peak assigned to O$_2$, which could result from the dissociation of CO$_2$ in the ion source of the instrument (Istomin et al., 1983). We emphasize that the uncertainty on the source of O$_2$ is specific to mass spectrometry (Newton, 1952).

The O$_2$ in situ detections have been dismissed as artifactual either because of the difficulties in reconciliation with the ground-based observations (Mills, 1999; Trauger and Lunine, 1983) or lack of known physical or chemical processes that could maintain 10s of ppm O$_2$ levels in the hot, reactive lower atmosphere of Venus (Krasnopolsky, 2006; Von Zahn et al., 1983). However, the multiple, consistent in situ detections would suggest that O$_2$ is indeed present at ~10s ppm.

The source of O$_2$ in the clouds of Venus is unknown and has been extensively discussed elsewhere (Bains et al., 2021a). One potential source of O$_2$ that was not considered by Bains et al. (2021a) is generation of O$_2$ during the chemical transformation of organics in the sulfuric acid cloud droplets, analogous to the speculative process first proposed by Hartley et al. (1989). Recently, the idea that there are organics in the clouds of Venus gained more traction with the proposed non-biological organic carbon cycle in the clouds (Spacek, 2021), opening the possibility for this chemistry to contribute to the in-cloud O$_2$ reservoir. However, the generation of $O_2$ from organic material reacting with concentrated sulfuric acid, if it happens at all, is thermodynamically unfavorable, requires many poly-carbonyls as intermediates (which are reactive and unstable in concentrated sulfuric acid (Bains et al., 2021b, 2021d)), and is likely too inefficient to account for all of the observed PVGC and VGC abundances of $O_2$.

---

[2] We note that the early high estimations of the abundance of O$_2$ in and below the clouds of Venus by Venera 4, Venera 5 and Venera 6 are likely erroneous due to the cross-reactivity of the chemical sensors with sulfuric acid. Recall that at the time of the early missions to Venus the composition of the clouds was completely unknown and the sensors have not been designed to account for high concentrations of sulfuric acid in the atmosphere (Florensky et al., 1978). Venera 11 and Venera 12 probes optical spectrometers have also provided upper limits of 50 ppm to the abundance of O$_2$ at altitudes below 60 km (Moroz, 1981).

Addressing the discrepancies between the measurements should be the main objective of future missions to Venus. Currently, it is difficult to reconcile the strong upper limits on the abundance of $O_2$ above 58 km (<3 ppm) imposed by ground-based observations with the in-cloud $O_2$ abundances detected by both Pioneer Venus and Venera probes (see, *e.g.*, Mills 1999). One expects to observe a gradient of $O_2$ from above to below the clouds if the presence of $O_2$ in the clouds is not spatially or temporally varied (which it might be as other atmospheric observations, for example unknown absorber, are spatially and temporally variable (see, *e.g.*, Lee et al., 2019 and Yamazaki et al., 2018). Such discrepancies can only ultimately be resolved by new in situ measurements of $O_2$ in the clouds of Venus.

**Phosphine ($PH_3$)**. The recent tentative detection of ppb levels of $PH_3$ in the atmosphere of Venus through millimeter-wavelength astronomical observations (Greaves et al., 2021b) is surprising as there is no known process capable of producing even a few ppb of $PH_3$ on Venus (Bains et al., 2022a, 2021c). Volcanically extruded phosphide minerals from the deep mantle have been recently proposed as a potential source of $PH_3$ (Truong and Lunine, 2021). However, phosphide-containing minerals, including those from deep mantle plume volcanic eruptions and meteoritic delivery, are an extremely unlikely source of ppb $PH_3$ on Venus (Bains et al., 2021c, 2022c, 2022d). For the former, a brand new mechanism for explosive volcanism would be required in addition to the fact that phosphide minerals easily oxidize during their transport to the surface (Bains et al., 2022c, 2022d). For the latter, Bains et al. (2021c) argued that the amounts of phosphides delivered by meteorites are too small to explain the observed abundance of $PH_3$, although Omran et al. (2021) provided a counter argument on the abundance of meteoritic delivery as a source of $PH_3$. Work of Bains et al. (2023a) showed that, regardless of the assumed value of $\Delta G°_{(g)}$ for $P_4O_6$, the formation of $PH_3$ from $P_4O_6$ in the Venusian atmosphere is thermodynamically unfavorable. A recent assessment of the photochemical production of $PH_3$ from $P_4O_6$ suggests a $PH_3$ upper limit of 2 ppb between 50 and 60 km (Wunderlich et al., 2023). The existence of $P_4O_6$ itself in the atmosphere of Venus, however, is uncertain (Bains et al., 2023a).

Since the initial $PH_3$ discovery was announced, several papers have questioned the detection, either on the grounds of data analysis (Akins et al., 2021; Snellen et al., 2020; Thompson, 2021; Villanueva et al., 2021) or an assignment of the observed millimeter wavelength absorption to mesospheric $SO_2$ rather than cloud-level $PH_3$ (Lincowski et al., 2021; Villanueva et al., 2021).

The authors of the original discovery have provided a response to the critiques, both on data processing and data interpretation (Greaves et al., 2021c, 2021a) and on arguing against $SO_2$ contamination (Greaves et al., 2022b). Although sulfur dioxide variability is significant even on day-to-day timescales in the mesosphere, $SO_2$ would need to have increased 10-fold planet-wide over only a few days for $SO_2$ to have mimicked $PH_3$ in the discovery data (Greaves et al., 2022b).

The in-cloud location of the $PH_3$ signal has also been debated. Lincowski et al. (2021) reported that the $PH_3$ must reside above the clouds to produce the 1.1 mm absorption, and independent re-analyses has confirmed this finding (Greaves et al., 2022b; Villanueva et al., 2021). If mesospheric, a phosphine interpretation of the 1.1 mm feature is challenging given the expected short lifetime (Bains et al., 2021c). Simultaneously, the near-contemporaneous $SO_2$ measurements by Greaves et al. (2022b) render $SO_2$ a challenging interpretation as well. Other possibilities include a yet unknown mesospheric replenishment mechanism for $PH_3$ (as $SO_2$ is replenished in the high atmosphere (Zhang et al., 2012)), or previously unknown absorption from another molecule, or the presence of a completely novel absorber altogether (Greaves et al., 2022b).

Several groups have used IR observations to provide strong upper limits above the clouds (in the low ppb to sub-ppb range) on the abundance of $PH_3$ (Cordiner et al., 2022; Encrenaz et al., 2020; Trompet et al., 2020). However, recent preliminary results suggest that the IR observational upper limits are not inconsistent with tentative phosphine detections if there is a difference in abundance between day and night (Greaves et al., 2022a).

See the work of Bains et al. (2022a) and more recently Cleland and Rimmer (2022) and Clements (2023) for a summary for the phosphine debate so far.

An independent re-analysis of the Pioneer Venus Neutral Gas Mass Spectrometer (LNMS) data (Mogul et al., 2021b) shows evidence of $PH_3$ in the clouds of Venus, via detection of unique $PH_3$ fragmentation ions. The re-analysis was for the in-cloud altitude of 51.3 km, and yields a $PH_3$ abundance of ~2 ppm.

The debate on the presence of $PH_3$ in the clouds of Venus continues and will likely only be resolved by in situ measurements of $PH_3$ gas in the Venus atmosphere. At the time of writing, $PH_3$ is being considered as one of the additional target gases for the upcoming DAVINCI mission (Queen et al., 2022).

**Methane ($CH_4$).** The low-mass volatile hydrocarbons methane ($CH_4$), ethane ($C_2H_6$), and benzene ($C_6H_6$) were detected in situ in the atmosphere of Venus by the LNMS on the Pioneer Venus Large Probe (Donahue and Hodges Jr, 1993; Mogul et al., 2021b). $CH_4$ in particular was measured to be present with an unexpectedly high abundance (1000–6000 ppm) in the lower atmosphere altitudes (Donahue and Hodges Jr, 1993). In contrast to other gases discussed in this section, the detection of $CH_4$ and other volatile hydrocarbons by the Pioneer Venus Large probe are likely an artifactual result due to an instrumental contamination and not a genuine atmospheric gas detection (Donahue and Hodges Jr, 1993). This interpretation has been recently bolstered by results of the reanalysis of the LNMS data (Mogul et al., 2022). The high abundance (1000–6000 ppm) of methane (as well as detection of benzene and other volatile hydrocarbons) below the clouds is likely a contamination from the spacecraft itself. It is not known whether the in-cloud detection of methane by LNMS is a contamination as well (Mogul et al., 2022). The PVGC did not detect $CH_4$, which placed upper calculated limits of atmospheric $CH_4$ at <10 ppm, <3 ppm, and <0.6 ppm at 51.6 km, 41.7 km, and 21.6 km, respectively (Oyama et al., 1980b). The Venera 14 Gas Chromatograph (VGC) did not detect methane and put an abundance upper limit at 0.5 ppm below 58 km (Mukhin et al., 1982). Remote observations with Earth-based telescopes put upper limits for $CH_4$ abundance in the lower atmosphere at < 0.1 ppm (Pollack et al., 1993). As with any unexplained detections, the presence of $CH_4$ in the clouds of Venus remains to be confirmed and reconciled with the established abundance upper limits.

Early ground-based observations also established abundance upper limits, above the clouds, for several volatile organics, including simple hydrocarbons, halocarbons, formaldehyde, other volatile carbonyls, and HCN (see (Moroz, 1981), their Table VI). Recently new upper limits on HCN and formaldehyde in the Venus lower-mesosphere have also been established using SOIR instrument on board Venus Express (Mahieux et al., 2023).

We note that the Venera 13 and Venera 14 mass spectrometers also detected peaks belonging to methane, as well as hydrocarbons. Those signals, however, have been interpreted as contamination (*i.e.*, "background peaks") and are not considered valid detections of Venusian atmospheric gases (Istomin et al., 1983).

**Ammonia ($NH_3$).** $NH_3$ is unexpected in an oxidized atmosphere. $NH_3$ has been tentatively detected by two separate probes. In 1972, the Venera 8 descent probe reported the presence of $NH_3$ in the lower atmosphere of Venus using bromphenol blue as an indicator of a basic atmospheric component (Surkov et al., 1973)[3]. The $NH_3$ measurement has been challenged as erroneous, due to the indicator's potential reactivity with sulfuric acid (Young, 1977). The Venera 8 detection of $NH_3$ was also discounted shortly after the measurement; Goettel and Lewis (1974) discarded it on the grounds of its unlikelihood in an atmosphere at thermodynamic equilibrium. The argument by Goettel and Lewis is now weakened as a growing list of gases in the atmosphere of Venus indicates thermodynamic disequilibrium (Esposito et al., 1997; Johnson and de Oliveira, 2019; Mogul et al., 2021b).

The recent re-assessment of the Pioneer Venus Large Probe Neutral Mass Spectrometer (LNMS) has also provided suggestive, although not conclusive, evidence for the presence of $NH_3$ in the Venus cloud layers (Mogul et al., 2021b).

While the chemical processes that may generate $NH_3$ in the Venusian clouds are unknown, assuming the tentative detections of the Venera probes and Pioneer Venus are correct, the possibility that $NH_3$ is a biological product remains (Bains et al., 2021a). $NH_3$ should be a prime target for measuring with new in situ probes due to its critical role in the potential habitability of the clouds (Bains et al., 2021a; Mogul et al., 2021a) (see Section 4). At the time of writing, $NH_3$ is being considered as one of the additional target gases for the upcoming DAVINCI mission (Queen et al., 2022).

Similarly to $O_2$, the ground-based observations impose an upper limit of 6 ppb on the $NH_3$ abundance above the clouds (Krasnopolsky, 2012a). Even stricter upper limits of 28.4 ppt on the abundance of $NH_3$ in the Venus lower-mesosphere have been established using SOIR instrument on board Venus Express (Mahieux et al., 2023). These upper limits for abundance above the clouds are difficult to reconcile with the tentative in situ observations, unless the $NH_3$ loss in the upper atmosphere is balanced by a constant production that is localized to the clouds and the stagnant haze layer below (Bains et al., 2021a).

---

[3] Venera 9 and Venera 10 descent module mass spectrometers reported in situ upper limits for $NH_3$ of <0.05% for altitude range of 37-45 km that are consistent with tentative Venera 8 detection (Surkov, 1977).

**Sulfur hexafluoride ($SF_6$).** $SF_6$ has been tentatively detected between 35 and 58 km altitude by Venera 14 GC, in an abundance of $0.2 \pm 0.1$ ppm (Mukhin et al., 1982).

$SF_6$ has the same GC retention time as $N_2O$, but $N_2O$ has been ruled out based on the unrealistically high abundance (~1%) required to explain the observed signal (Mukhin et al., 1982), such high abundance is also inconsistent with PVGC upper limits for $N_2O$ in the atmosphere of Venus, <200 ppm, <70 ppm, and <10 ppm at 51.6 km, 41.7 km, and 21.6 km, respectively (Oyama et al., 1980b).

It is also unlikely that the tentative detection of $SF_6$ resulted from the contamination from the spacecraft. $SF_6$ is a commercial electrical insulator in high-voltage switches and transformers used in electrical gear operating at tens of kilovolts (Simmonds et al., 2020), and it is very unlikely that this engineering solution would be used in a spacecraft.

On Earth, trace amounts of natural $SF_6$ exist in volcanic rocks in rift zones, faults, igneous intrusions, geothermic areas, and diagenetic fluids (Busenberg and Plummer, 2000). $SF_6$ is predominantly present in fluorites and some granites, while basalts, for example, do not contain detectable $SF_6$ (Busenberg and Plummer, 2000; Harnisch and Eisenhauer, 1998). The exact process of production of natural $SF_6$ on Earth is unknown. It is also unclear whether $SF_6$ is directly made by volcanic processes on Earth or is $SF_6$ only released in association with volcanic activity. The work of Harnisch and Eisenhauer studied the gases from several volcanic fumaroles, for example, from Sicilian (Vulcano Island and Etna) and Japanese (Satsuma Iwojima and Kuju) volcanoes, and found that they are not significant sources of $SF_6$ (Harnisch and Eisenhauer, 1998). However, the underlying rocks of these volcanoes are not granitic and as a result might lack the source for $SF_6$ (Harnisch and Eisenhauer, 1998). The pre-industrial atmospheric equilibrium concentration of $SF_6$ on Earth is estimated to be <0.06 ppt (Busenberg and Plummer, 2000). See also the recent work of (Seager et al., 2023a) for detailed discussion of the possible planetary sources of $SF_6$.

Since Venus is significantly H-depleted, one would expect that it would have a different profile of F-containing volatiles erupted by volcanoes than that of Earth. On Earth, HF is the main source of volcanic F (see, *e.g.*, Cheng, 2018). On Venus, HF is also volcanic, but it is likely that the majority of F is erupted as other compounds, for example $SSF_2$, $COF_2$, $FClCO$, and $SOF_2$, etc. (Zolotov and Matsui, 2002). Therefore, it is not unexpected that, in an H-depleted environment of Venus, with abundant sulfur, $SF_6$ could also be a volcanic product released in significantly higher abundance than on Earth. $SF_6$ could also be the result of weathering of fluorite minerals abundance of which on Venus is poorly constrained.

*3.5. Unexplained Gas Vertical Abundance Profiles of $SO_2$ and $H_2O$*

The atmospheric vertical abundance profiles of sulfur dioxide ($SO_2$) in the Venus cloud layers and $H_2O$ in and above the clouds remain unexplained. The presence of $SO_2$ is expected in the atmosphere of Venus and in the clouds. $SO_2$ is a common volcanically produced gas. However, the observed abundance of $SO_2$ ascending through the Venus cloud layers drops from an average of ~150 ppm below the clouds to sub-ppm levels above the clouds. This depletion cannot be currently explained by known atmospheric chemistry (see Rimmer et al., 2021)). There is, therefore, missing atmospheric chemistry of some kind, a fact that has been recognized by Venus researchers for decades (see, *e.g.*, Bierson and Zhang, 2019; Marcq et al., (2018); Mills et al. (2007); and Vandaele et al. (2017)).

The $SO_2$ depletion in the clouds is unlikely to be solely caused biologically by the three sulfur-energy metabolic strategies postulated previously (Schulze-Makuch et al., 2004; Schulze-Makuch and Irwin, 2006), which have been recently investigated by Jordan et al. (2022). Sulfur energy metabolisms investigated by these authors require reduced species as input, either hydrogen-containing compounds ($H_2S$ or $H_2$) or carbon monoxide (CO). Reduced compounds are expected to be rare in Venus' oxidized atmosphere (see, *e.g.*, Marcq et al., 2018), and so it is not surprising that these metabolic strategies could not solely explain the $SO_2$ depletion. Therefore, either other sulfur metabolic strategies are at play or the $SO_2$-depletion in the clouds has another explanation altogether, including the possibility of multiple contributing processes acting at once.

The observed abundance of water vapor ($H_2O$) above the clouds also does not match the $H_2O$ abundance profile predicted by atmospheric photochemistry models (Bierson and Zhang, 2019; Greaves et al., 2021b; Winick and Stewart, 1980). As in the $SO_2$ case, additional unknown atmospheric chemistry is needed to explain the observations of $H_2O$. We discuss the possible explanation for the $SO_2$ and $H_2O$ abundance profiles in Section 4.

We would be remiss not to emphasize the extreme aridity of the Venus cloud environment as a significant challenge to life as we know it. In extremely dry environments, terrestrial life can survive as spores or other inactive forms but would not be actively growing and, therefore, unable to support a sustainable biosphere. Even under the assumption that life resides inside cloud particles, the water activity is extremely low, because any water molecules inside the particle will be tightly bound to sulfuric acid.

The extreme dryness of Venus' atmosphere has been considered a well-known fact for decades (see, *e.g.*, Donahue and Hodges Jr, 1992), having been described on many occasions (see, *e.g.*, Bains et al., (2021c); Bains et al., (2021d); and Seager et al., 2021)). However, there appears to be a great variability in observed water vapor abundance values. Repeated measurements by in situ probes vary from 5 ppm to 0.2% (reviewed by Rimmer et al. (2021) (see also Table 4). In situ measurements of water from the 1970s and 1980s Soviet VeGa and Venera probes and NASA Pioneer Venus give average abundances of 200–2000 ppm in the middle/lower clouds (58–48 km) and 5000 ppm just below the clouds (41.7 km) (Table 4). These values are considerably higher than the global average of around 30 ppm, which comes from water vapor abundances derived from the spectrometric measurements by Venera probes (Ignatiev et al., 1997), the Galileo spacecraft (Drossart et al., 1993), Venus Express orbiter (Bézard et al., 2009), and supported by ground-based near IR spectroscopy (De Bergh et al., 1995) (with a notable exception of the tentative 200 ppm value measured by Bell et al. (1991) near 2.3 μm). Ignatiev et al. (1997) dismissed the observations based on "contact methods" (CM) in favor of spectroscopic methods. They discarded the values without being able to "point to specific shortcomings of contact methods," because the values are so much higher than spectroscopic observations (Ignatiev et al., 1997). Recently, Mogul et al. (2021a) suggested that the higher values of the "contact methods" are due to water from the cloud particles. If such an interpretation is true, it would mean the cloud particles have more water than our current understanding of the cloud particles allows.

Nevertheless, the highest values, if confirmed, indicate the presence of local "habitable" regions with higher-than-average humidity. While all global Venus atmosphere models may, therefore, represent an average of extremely arid "desert" regions, there may exist some localized, more humid regions (albeit still far drier than any environment on Earth) future Venus mission planners would endeavor to remeasure H$_2$O content of the clouds in multiple locations.

The very low water activity is not a definitive refutation of the possibility of life in the clouds of Venus (Bains et al., 2023b). Life may have a completely different biochemistry to that of Earth, for example one based on concentrated sulfuric acid instead of water as a solvent (Bains et al., 2021d; Seager et al., 2023b). Or life on Venus, if it exists, may have evolutionary adaptations without precedent here on Earth to actively extract water from the dry atmosphere or from water tightly bonded to sulfuric acid inside the Venus cloud particles.

**Table 4.** Summary of controversial atmospheric observations and unexplained cloud property measurements. VGC is Venera Gas Chromatograph; CM is Contact Methods; PVGC is Pioneer Venus Gas Chromatograph; LNMS is Pioneer Venus Large Probe Neutral Gas Mass Spectrometer; VMS is Venera Mass Spectrometer; XRF is X-ray Fluorescence Spectrometer; LCPS is Pioneer Venus Large Cloud Particle Size Spectrometer.

| Venus Atmospheric Observable | In Situ Method | Controversy | Comments |
|---|---|---|---|
| $H_2O$ | VGC; CM | Highly variable measurements that are inconsistent with each other and the spectroscopic methods. "Contact methods" (CM) give generally much higher $H_2O$ abundances than other in situ methods. | CM methods give $H_2O$ abundances in a range of thousands of ppm. For example, VeGa estimates 1000 ppm $H_2O$ at 50-60 km, decreasing to 150 ppm at 25-30 km (Surkov et al., 1987). Venera 14 humidity sensor gave the value of 2000 ppm at 46-50 km (Surkov et al., 1983, 1982) while Venera 14 GC gave 700 ppm at 49-58 km (Mukhin et al., 1982). PVGC also suggested high $H_2O$ water mixing ratios <600 ppm at 51.6 km, ~5000 ppm at 41.7 km and 1350 ppm at 21.6 km (Oyama et al., 1980b). Venera 12 GC does not agree with such a high $H_2O$ abundances and provides upper limits of 200 ppm below 42 km (Gelman et al., 1979b). Ultimately such discrepancies can only be resolved by new in situ measurements of water abundance at multiple locations in the atmosphere and inside the cloud particles. |
| $O_2$ | PVGC; VGC | Ground-based observations provide upper limits for the abundance of $O_2$ above the clouds (Mills, 1999; Trauger and Lunine, 1983) that are inconsistent with the abundance reported by Venera and Pioneer probes. | There is no apparent reason to consider ppm levels of $O_2$ detected by PVGC and VGC as erroneous. Lab studies with PVGC confirm that $O_2$ is not the product of thermal decomposition of the $SO_3$ gas or $H_2SO_4$ and that the detection is robust (Oyama et al., 1979a). |
| | LNMS; VMS | The detections of $O_2$ by MS are considered unreliable. $O_2$ ions could originate from reaction of $CO_2$ in the mass spectrometer itself. | MS detections of $O_2$ are uncertain due to overlapping mass (isobaric species) and the potential formation of $O_2$ in the instrument itself. |
| $NH_3$ | Chemical Sensor | Ground-based observations (Krasnopolsky, 2012a) and Venus Express upper limits (Mahieux et al., 2023) are difficult to reconcile with the in situ measurements. | The bromophenol blue chemical sensor was used as an indicator of a basic atmospheric component (Surkov et al., 1977). The change of color was registered by photoresistors. The results of the measurement are tentative (Surkov et al., 1974, 1973) and could be a result of a false positive detection due to cross-reactivity with sulfuric acid vapor (see also discussion in (Bains et al., 2021a)). |
| | LNMS | The detection of $NH_3$ is tentative due to possible isobaric species. | $NH_3$ identification the re-analyzed LNMS data remains tentative. |
| $PH_3$ | LNMS | $P^+$ ion identified in the LNMS data is sufficiently separated and appears to be real (Mogul et al., 2021b). The identity of the parent gaseous compound of the $P^+$ ion is unknown. | The detection of $P^+$ ion is robust. $PH_3$ is the simplest gas that fits the data (Mogul et al., 2021b). Other known volatile P-species at the cloud level temperatures do not provide a good fit to the LNMS data. |
| $H_2S$ | VGC | The abundance of $H_2S$ is uncertain. Ground-based observations put upper limits of <23 ppb above the cloud tops (Krasnopolsky, 2008). The VGC detection appears to be inconsistent with PVGC upper limits: <40 ppm at 51.6 km, <10 ppm at 41.7 km and <2 ppm at 21.6 km (Oyama et al., 1980b). | $H_2S$ has been first tentatively detected by Venera 11 and Venera 12 GC although without a firm constrain on the abundance. Venera 14 GC detection remains much more robust than earlier measurements but significant uncertainty on the abundance of $H_2S$ remains (Mukhin et al., 1982). Possible discrepancy between Venera 14 GC measurement at 29-37 km attitude and PVGC upper limits remains. |
| | LNMS | Tentative identification of $H_2S$ in the re-analyzed LNMS data confirms the original LNMS detection; significant overlap of isobaric species (Mogul et al., 2021b). Abundance of $H_2S$ is uncertain. | The reanalyzed data from LNMS indicate the presence of $H_2S$ at 51 km (Mogul et al., 2021b). Earlier LNMS inference of $H_2S$ abundance (from ratio to $^{36}Ar$) suggest 3 ppm ± 2 ppm below 24 km (Hoffman et al., 1980a). |
| HCN | LNMS | Only tentative detection in re-analyzed LNMS data (Mogul et al., 2021b). The early ground-based observations propose | HCN is thermodynamically disfavored in the atmosphere of Venus and is reactive to conc. |

| | | 1 ppm as the upper limits on HCN abundance above the clouds (see (Moroz, 1981), their Table VI). See also new upper limits on HCN in the Venus lower-mesosphere from Venus Express (Mahieux et al., 2023). | sulfuric acid. If confirmed HCN could be an important element of the Venusian nitrogen cycle. |
|---|---|---|---|
| $NO_x$ | LNMS | Tentative identification of $NO_x$ in the re-analyzed LNMS data (Mogul et al., 2021b). See also early ground-based observations upper limits on $NO_x$ abundance above the clouds (see (Moroz, 1981), their Table VI) and the new upper limits on $NO_x$ in the Venus lower-mesosphere from Venus Express (Mahieux et al., 2023). | LNMS is the only in situ instrument that detected $NO_x$. PVGC provided upper limits for $N_2O$ in the atmosphere of Venus (see (Oyama et al., 1980b), their Table 3) but in situ upper limits for other nitrogen oxide species are unknown. |
| $CH_4$ | LNMS; VMS | $CH_4$ detected by LNMS is likely an artefact arising from the outgassing within the spacecraft itself (Donahue and Hodges Jr, 1993; Mogul et al., 2022). | LMNS detection of $CH_4$ is considered artefactual (Donahue and Hodges Jr, 1993; Mogul et al., 2022), VMS detection is a background detection (Istomin et al., 1983). PVGC measurements provided strong upper limits to $CH_4$, $C_2H_4$, $C_2H_6$ and $C_3H_8$ abundance (see (Oyama et al., 1980b), their Table 3). |
| complex organic molecules | Never attempted | No uncontested evidence of organic molecules in the atmosphere of Venus. | The direct in situ detection of organic chemicals in the atmosphere of Venus has never been attempted. The potential for organic carbon cycle in the atmosphere of Venus exists (Spacek, 2021). Future Venus missions should aim to identify organic molecules in the cloud particles. |
| non-volatiles (e.g. P and Fe) | XRF | No reliable estimates on abundances of non-volatile species, in particular P, can be derived from VeGa 1 and VeGa 2 detections (Krasnopolsky, 1989). | The detection of P-bearing species was confirmed by the Pioneer Venus LNMS re-analysis (Mogul et al., 2021b). The debate on the possibility of the phosphoric acid as a significant component of the cloud aerosols continues and is an intriguing area for future investigation (Milojevic et al., 2021). The detection of P in the clouds is unexpected and should be confirmed by future missions. The detection of Fe (Petrianov et al., 1981) by Venera 13 and Venera 14 XRF is less controversial and generally accepted as valid (e.g. (Krasnopolsky, 1989)). |
| mode 3 particles | LCPS | The existence of the Mode 3 particles has been questioned (e.g. (Toon et al., 1984)) and is not supported by Venera nephelometer measurements although Venera measurements provided less conclusive data on the modality of particles than Pioneer Venus. | The reanalysis of the LCPS data reaffirmed the existence of the Mode 3 particles although their solid, crystalline nature is uncertain and can only be confirmed with new in situ measurements (Knollenberg, 1984). |

## 4. Challenging the General Consensus of Venus Cloud Composition and Acidity

The Venus clouds' main constituent is particles composed of concentrated sulfuric acid droplets (see, *e.g.*, Knollenberg et al., 1980; Moshkin et al., 1986) (Table 2). This paradigm is supported by several findings.

1] Photochemical models of the atmosphere are consistent with $H_2SO_4$ clouds. The models predict $H_2O$, $SO_3$, and $H_2SO_4$ to be present throughout the atmosphere (see, *e.g.*, Bierson and Zhang, 2019) and gaseous $H_2SO_4$ (Oschlisniok et al., 2021, 2012), as well as gaseous $H_2O$ and $SO_2$ (reviewed by Rimmer et al., (2021) are measured throughout. Simple, condensation models for the $H_2SO_4$-$H_2O$ gas-cloud system on Venus also track the fate of liquid Mode 2 particles and confirm that they are mostly composed of $H_2SO_4$ (Dai et al. 2022).

The consensus model is that formation of clouds on Venus is photochemically driven (see, e.g., Krasnopolsky (2012b, 2007). Sulfuric acid vapor is first made at > 70 km.

$CO_2 + h\nu \rightarrow CO + O$

$SO_2 + O + M \rightarrow SO_3 + M$

$SO_3 + 2\ H_2O \rightarrow H_2SO_4 + H_2O$

The $H_2SO_4$ vapor condenses out, creates the droplets, and as the droplets rain down, sulfuric acid thermally dissociates in the lower atmosphere (below 40 km) (Krasnopolsky, 2013, 2007). A fraction of $H_2SO_4$ also likely reforms from the $H_2O$ and $SO_3$ near the bottom of the clouds (Krasnopolsky, 2007). The measured and modeled levels of $H_2O$, which together with $SO_3$ will efficiently form $H_2SO_4$, support the theory that the clouds of Venus contain sulfuric acid (Krasnopolsky, 2007; Oyama et al., 1980b; Vinogradov et al., 1970).

2] Gaseous $H_2SO_4$ has been detected and measured by microwave spectrometry, supporting the photochemical concept of $H_2SO_4$ cloud formation (Oschlisniok et al., 2021, 2012).

3] The proposed interpretation of the inferred refractive index of the cloud droplets is that the clouds are made of at least 70% w/w sulfuric acid and less than 30% w/w water (Palmer and Williams, 1975; Young, 1973). The concentration of sulfuric acid in droplets is derived through modeling (e.g., James et al., 1997) of light scattering to match in situ data. The concentration of $H_2SO_4$ is lower in the top clouds and increases towards the bottom of the clouds as the temperature increases (summarized in Table 2; following information from the work of Titov et al. (2018) Table 1). Furthermore, the concentrated solution of sulfuric acid ($H_2SO_4$-$H_2O$) has been found to be in good agreement with ground-based polarization data (Hansen and Hovenier, 1974) before the in situ probes.

4] The VeGa chromatographic measurements of the cloud aerosols are the only dedicated in situ estimates of the sulfuric acid concentration in the cloud particles[4]. The Vega measurements confirmed that the clouds are primarily composed of concentrated sulfuric acid and water (Gelman et al., 1986; Porshnev et al., 1988, 1987). VeGa chromatograph collected cloud aerosols on the carbon fibers between the altitudes of 63 km to 48 km. Gases—$SO_2$, $H_2O$, and $CO_2$—evolved upon heating of the collected sample on a carbon substrate are consistent with sulfuric acid droplets (Gelman et al., 1986)[5]. We note, however, that Gelman et al. (1986) also suggested that the cloud layers may consist of particles of more complex composition than a pure aqueous sulfuric acid solution (Gelman et al., 1986). This suggestion was later supported by the preliminary calibration experiments of the VeGa gas chromatograph (Mukhin et al., 1987). The experiments results show that the pyrolyzed sulfuric acid aerosols evolved significant amounts of $H_2S$, which suggests other unknown condensed or dissolved constituents of the aerosols beyond the sulfuric acid and water (Mukhin et al., 1987).

The "average concentration" of sulfuric acid in the cloud droplets is ~85% w/w $H_2SO_4$ (Titov et al., 2018). However, the concentration of sulfuric acid across the cloud deck likely varies significantly (Krasnopolsky, 2015). The concentration reaches ~70% in the top clouds, while in the lower clouds the concentration could reach >100%, that is, "fuming" sulfuric acid or oleum ($H_2S_2O_7$, or a solution of $SO_3$ in $H_2SO_4$) (Titov et al., 2018). The measured cloud particle refractive index suggests that the chemical composition of the clouds may include a number of chemicals that may be mixed with concentrated $H_2SO_4$ or be completely different from $H_2SO_4$ (Knollenberg et al., 1980; Ragent and Blamont, 1979). For example, the droplet sulfuric acid concentration could be highly variable, between 30% and >100% (Section 3.3).

Recently, Rimmer et al. (2021) proposed a photochemical model of the atmosphere of Venus that also includes a new view of the cloud chemistry. The model postulates that the cloud droplets are not homogenous in composition and a fraction of the cloud particles are neutralized solid or semi-solid salt particles, instead of liquid concentrated sulfuric acid droplets.

A base is needed to neutralize (convert to salts) all the sulfuric acid in a cloud particle, so that the pH of the particle reaches >0. When pH of the liquid in the droplet is >0, the equilibrium between $SO_2$ and sulfite ($HSO_3^-$) is pulled towards the sulfite, thus removing $SO_2$ from gas phase, that is, trapping $SO_2$ in the droplet as sulfite salts, providing a mechanism that could explain the mysterious depletion of $SO_2$ in the atmospheric cloud layers, and the vertical abundance profile of $H_2O$ in and above the clouds (see, *e.g.*, Rimmer et al. 2021) for details on the model).

The identity of the putative acid-neutralizing base is unknown. Bains et al. (2021a) postulated that biologically produced $NH_3$ could be a neutralizing base for the Venusian cloud droplets.

The Bains et al. (2021a) model calculates the amount of $NH_3$ needed to neutralize all the sulfuric acid in the droplet and the amount needed to trap $SO_2$ as ammonium sulfite, therefore explaining the unusual $SO_2$ abundance profile (Bains et al., 2021a). In the model, all the sulfuric acid in the Mode 3 particles is reacted with $NH_3$ to form ammonium sulfate salts. The particles are, therefore, either solid or a slurry of solid and fluid. For the particles to absorb $SO_2$ they must have a pH>0 if fluid is present. Whether fluid is present, and its exact pH, will depend on the water activity in the clouds, which is poorly constrained (Section 3.5).

---

[4] The Pioneer Venus LNMS was not designed to sample aerosols, however the gases evolved after the gas inlet blockade was lifted are consistent with droplets composed of 85% w/w $H_2SO_4$ and 15% w/w $H_2O$ (Hoffman et al., 1980b).

[5] Note that if organic carbon compounds were present in the droplets, these would not have been identified over the background of $CO_2$ evolved from the carbon fibre filters.

We note that the neutralization of the sulfuric acid droplets by $NH_3$ in Venus' atmosphere could proceed analogously to the neutralization of sulfuric acid aerosols by $NH_3$ in Earth's stratosphere (see, *e.g.*, Huntzicker et al., 1980), including neutralization of particles of up to 80% acid.

The removal of concentrated sulfuric acid in the droplet by reacting it with $NH_3$ has a crucial outcome for the overall habitability of the clouds. The model postulates that some of the cloud particles are much less acidic than previously thought, with a pH between -1 and 1 (Rimmer et al., 2021), instead of an acidity of approximately -11 (on the Hammett acidity scale), which is uninhabitable for terrestrial life (Seager et al., 2021).

The amount of $NH_3$ base modeled by Bains et al. (2021a) explains many of the Venus' lingering atmospheric chemical anomalies (Table 5). The model agrees with the tentative detections of $NH_3$ in the clouds and below the clouds. In agreement with observations, the model predicts that both $SO_2$ and $H_2O$ will be present in and above the clouds but at substantially lower abundance than they are below the clouds. The model also provides the explanation for the in-cloud abundance of $O_2$ and the presence of $NO_x$ and $H_2S$ in the atmosphere. Finally, independent of atmospheric chemistry, the predictions of Bains et al. (2021a) on the Mode 3 cloud particle composition have been supported by the re-analysis of the Pioneer Venus legacy data on the refractive index of the Venusian cloud droplets (Mogul et al., 2021a). The re-analyzed data on the refractive index of cloud particles also suggest ammonium hydrogen sulfate ($NH_4HSO_4$) salts as components of Mode 3 particles (Mogul et al., 2021a).

We note that one base that was not considered by Bains et al. (2021a) is hydroxylamine ($NH_2OH$). We find that formation of hydroxylamine as an $H_2SO_4$ neutralizing agent is thermodynamically less costly (requires less energy) than formation of $NH_3$. As in the case of $NH_3$, production of $NH_2OH$ from $N_2$ requires a release of an oxidized product, and, as in the case of $NH_3$, the most energy- and water-efficient oxidized product is $O_2$ (see Supplementary Information, Table S1 and Table S2). However, hydroxylamine reacts readily with aldehydes and ketones and cleaves some peptide bonds (Bornstein and Balian, 1977), which means it is unlikely to be tolerated in high concentration by organisms with an Earth-like biochemistry. Interestingly, hydroxylamine itself reacts with sulfur dioxide to form sulfamic acid (Greenwood and Earnshaw, 2012), hinting at another potential chemical route to removing $SO_2$ from the clouds through biological action.

In situ Venus atmosphere measurements (Table 6) can verify the acidity and composition of the Venusian particles (Kaasik et al., 2022). A dedicated mission could confirm the non-spherical, semisolid nature of Mode 3 cloud particles, identify them as ammonia salts, and measure the acidity of the cloud particles, especially Mode 3 cloud particles. The confirmation of the decades-long anomalous gas abundances should also be a priority of any astrobiology-focused mission, as well as any Venus mission that focuses on the chemistry of the clouds and the atmosphere (see, *e.g.*, Agrawal et al., 2022; Buchanan et al., 2022; Seager et al., 2022a). Such measurements should include confirmation of the existence of $O_2$ and $NH_3$ as well as other nitrogen species, $NO_x$ for example, that could be indicators of an active nitrogen cycle in the atmosphere (see Figure S1 and the Supplementary Information).

We emphasize that, while our interest in Venus is motivated by astrobiology, all of the above measurements will have value regardless of what they find. As discussed, there are many unexplained aspects of Venus' atmospheric and cloud chemistry and resolving them will be of value regardless of whether the resolution involves the discovery of life. For example, the model of Bains et al. (2021a) postulated that atmospheric abundance of $O_2$ and $NH_3$ reported by in situ measurements are real and the result of biological neutralization of the Mode 3 droplets. If accurate measurements show that Mode 3 particles have a pH of around 0, but that there is no $NH_3$ or $O_2$ present in the atmosphere, that would rule out Bains et al.'s mechanism and suggest that the abiotic mineral-based mechanism for $SO_2$ removal suggested by Rimmer et al. (2021) could be at play. Even if none of the unexplained observations reported here are confirmed by accurate measurement, that in itself would resolve over forty years' of uncertainty and confirm the "null hypothesis" that the consensus model of Venus' atmosphere is correct. Thus, while Table 6 is cast in terms of its astrobiological significance, the measurements suggested in Table 6 will be of value regardless of the outcome.

**Table 5.** Venusian atmospheric characteristics explained by the presence of $NH_3$ base in the atmosphere of Venus (Bains et al., 2021a).

| Observable | Model Challenge[a] | Model Prediction[b] |
|---|---|---|
| $NH_3$ | High fidelity altitude abundance profile measurements do not confirm tentative detections. Detailed assessment of possible sources and sinks for $NH_3$ to confirm that the expected abundance of $NH_3$ is sufficient to act as a neutralizing agent. | The presence of $NH_3$ in the clouds and below the clouds is consistent with tentative detections of $NH_3$ (Table 3). Presence of $NH_3$ in clouds would foster production of salts and result in non-spherical Mode 3 particles. |
| $SO_2$ | Models that only include H from $H_2O$ and low HCl abundances in clouds do not explain vertical depletion above the clouds (>70 km). | $NH_3$ provides a mechanism that explains the depletion of $SO_2$ in the atmospheric cloud layers. $NH_3$ provides additional H budget that impacts $SO_2$ budget, replicating observed $SO_2$ depletion. |
| $H_2O$ | Models that only include H from $H_2O$ and low HCl abundances in clouds do not explain vertical depletion above the clouds (>70 km). | $NH_3$ explains the vertical abundance profile of $H_2O$ in and above the clouds. $NH_3$ provides additional H budget that impacts $H_2O$ budget, replicating observed levels. |
| $O_2$ | High fidelity altitude abundance profile measurements exclude the possibility of the coexistence of $NH_3$ and $O_2$ in the cloud layers. | If the chemistry of $NH_3$ production is the source of $O_2$, then the model predicts an order of 1 ppm $O_2$ in the cloud level of 50—60 km. |
| $H_2S$ | High fidelity altitude abundance profile measurements do not confirm $H_2S$ in the haze layer below the clouds. | If $NH_3$ is present in the Venus atmosphere, $H_2S$ is a result of disproportionation of $NH_4HSO_3$ that yields $NH_3$, $H_2S$, and $H_2O$ to the atmosphere below the clouds, and hence is a unique output of the Bains et al. model (Bains et al., 2021a). |
| $NO_x$ | No $NO_x$ detected within or below the clouds, insufficient data on oxidizing processes at Venus formation and in present day atmosphere to assess the expected abundance of $NH_3$. | If $NH_3$ is present in the atmosphere then it is oxidized to $NO_x$ within and below the clouds. |
| mode 3 particles | The search for $NH_4^+$ salt ions within the cloud particles gives negative result, meaning no neutralization of acid happens, or other salt ions are detected that could neutralize the acid and act as a base instead of $NH_4^+$ (e.g. $Ca^{2+}$ coming from hydroxide minerals from the surface (Rimmer et al., 2021)). | If $NH_3$ is the main neutralizing agent of the sulfuric acid cloud droplets, then the Mode 3 cloud particles in the lower clouds must be solid supersaturated in ammonium salts, with a small liquid phase, and therefore are not liquid, spherical droplets of concentrated sulfuric acid. This view is supported by both (Bains et al., 2021a) and (Mogul et al., 2021a). |
| stagnant haze layer (31—47 km) | The chemical composition of the stagnant haze layer does not match the model's predictions (small dry salt particles and coexistence of gases like $NH_3$, $H_2S$, $O_2$, $SO_2$, $NO_x$, $N_2$). | Thermal disproportionation of the salts generates gas that shatters the particles at the cloud base, the fragmented particles form the haze. |

[a] Model Challenge: what potential future observations, measurements and results would falsify or challenge the Bains et al. model (Bains et al., 2021a).
[b] Model Prediction: predictions of the Bains et al. model (Bains et al., 2021a).

## 5. Summary and Conclusions

We have described a number of observed Venus atmosphere and cloud properties that have not been previously explained or explored by Venus chemical or planetary evolution models. These shortcomings of the available models are direct evidence of our gaps in understanding of Venus due to insufficient data (both in situ at Venus and lab-based), which has led to many open questions about the mechanics of Venus's atmosphere (Gillmann et al., 2022; Marcq et al., 2018; Mills et al., 2007; Way and Del Genio, 2020).

Significant uncertainties remain embedded into the emergent consensus model of Venusian photochemistry. For example, at high altitudes (>80 km), measurements report the presence of much higher concentrations of $SO_2$ than predicted by models beforehand, termed the "$SO_2$ inversion layer" (Belyaev et al., 2012; Sandor et al., 2010; Zhang et al., 2012). This inversion layer is unexpected because $SO_2$ should be readily photo destroyed at such high altitudes (Mills, 1998; Yung and Demore, 1982; Zhang et al., 2010). Photolysis of supersaturated sulfuric acid or $S_8$ aerosols is a possible explanation for the inversion layer and is accessible to empirical test via laboratory experiment and observational confirmation (Zhang et al., 2012). However, confirmation has not yet been obtained such that the upper-atmosphere $SO_2$ cycle remains in doubt.

We conclude this paper with a call for repeated observations of the Venusian atmosphere mysteries with modern instrumentation and for further re-analysis of the legacy data. It is clear that there are a lot of unknowns about Venus. Repeated, high-fidelity, in-situ observation of atmosphere and cloud properties should be a paramount objective of future missions to Venus, as the presence of unexplained chemicals in the atmosphere might be tied to the habitability of the clouds, biological activity, or unknown chemistry.

New NASA missions to Venus such as DAVINCI (Garvin et al., 2022) and VERITAS (Freeman et al., 2016) and ESA missions to Venus such as EnVision (de Oliveira et al., 2018)) will add data to resolve some of the lingering questions about the planet, but none of the planned missions are equipped to directly sample and analyze the chemical composition of cloud particles. Further, additional measurements are needed to characterize atmospheric gases. Therefore, there remains a critical opportunity to directly sample and analyze the essential properties of the Venusian clouds. Unexplained chemical anomalies, including the possible presence of $NH_3$, tens of ppm $O_2$, the $SO_2$ and $H_2O$ vertical abundance profiles, and the unknown composition of Mode 3 particles, have lingered for decades, and their resolution might reveal unknown chemistry that is, in itself, worth exploring even in the absence of life.

The habitability of the Venusian clouds should also be explored by new in situ missions (Seager et al., 2022a). The acidity of the Venus cloud droplets has not been measured directly and could be key to cloud particle habitability. Similarly, no previous mission has directly searched for organic chemistry in the cloud particles. Rocket Lab Mission to Venus is planned for launch in January 2025 (French et al., 2022). The Rocket Lab mission will carry a probe containing the Autofluorescence Nephelometer (AFN) to search for autofluorescence indicative of organic molecules in cloud particles (Baumgardner et al., 2022). Detection of organic molecules, if found predominantly in the larger particles, would be an indicator of life.

New missions should aim to address each of the above objectives and continue where the pioneering missions from nearly four decades ago left off.

In the meantime, a public release of original data from the Soviet Venera and VeGa missions, as has been done recently for Pioneer Venus LNMS data (Mogul et al., 2021b), could enable further support or refutation of current models and predictions and would provide needed context for future mission results.

**Table 6.** Venusian atmospheric observations, their astrobiological context, required future measurements and possible mission science outcomes. See also (Seager et al., 2022a) for the detailed discussion of the atmospheric observables in the context of the science objectives and mission outcomes of the planned missions to Venus. Table modified from (Seager et al., 2022a) under CC BY 4.0 license.

| Observable | Astrobiological Motivation or Hypothesis | Required Measurements | Mission Science Outcomes |
|---|---|---|---|
| $NH_3$ | Indicator of habitability of the clouds (potential "neutralizing agent" of cloud droplets). Indicates an unknown chemical process contributing to the planetary nitrogen cycle. Challenges the notion that the clouds are solely composed of liquid droplets of concentrated sulfuric acid. | Measure altitude-dependent abundance profile of gaseous $NH_3$ within the clouds and below to 1 ppb precision combined with the search for $NH_4^+$ salt ions within the cloud particles. Measurements done at several latitudes, day vs night, by several probes would distinguish between localized vs global distribution of $NH_3$ and inform sources and sinks. | _Detection_: The abundance vs altitude profile constraints the source of $NH_3$ and tests the validity of the models and their implications. _Non-detection_: The $NO_x$ species (if confirmed) could not be the result of oxidation of $NH_3$; Reconciles the upper limits provided by the remote observations with the tentative in situ detections; Puts clear constraints on the chemistry of the cloud droplets and on the chemical processes in the atmosphere. |
| $SO_2$ | Variable profile, including in-cloud depletion, indicative of unknown chemistry in the atmosphere. | Measure altitude-dependent abundance profile of gaseous $SO_2$ measured from above the clouds to below the clouds to 1 ppb precision to characterize the degree of depletion of $SO_2$ within the clouds. | _Detection_: The abundance vs altitude profile constraints the source of $SO_2$ and other $SO_x$ gases and tests the validity of the models and their implications; Puts clear constraints on the chemical processes in the cloud droplets and the atmosphere. |
| $H_2O$ | The amount of water in the clouds is not uniform and is | Measure altitude-dependent abundance profile of water | _Detection of anomalously high abundance values_: |
| | locally variable. High abundance of $H_2O$ is an indicator of relatively greater habitability of local regions within the clouds. | vapor to 1 ppb precision together with the measurement of the water content of the cloud particles. Measurements done at several latitudes, day vs night, by several probes would distinguish between localized vs global distribution of $H_2O$ and inform sources and sinks. | Confirmation that the amount of water in the clouds is not uniform and is locally variable. Variable profile would be indicative of unknown cloud particle chemistry. *No anomalously high values detected*: Reconciles the values and upper limits provided by the remote and in situ spectroscopic observations with the tentative in situ detections. |
| $O_2$ | Potential sign of life or unknown abiotic chemical processes in the clouds. | Measure altitude-dependent abundance profile of $O_2$ to 1 ppb precision especially in the clouds and below. Measurements done at several latitudes, by several probes would distinguish between localized vs global distribution of $O_2$ and inform sources and sinks. Establishing the co-existence of $NH_3$ and $O_2$ in the cloud layers tests the hypothesis of biological production of both gases. | *Detection*: The abundance vs altitude profile constraints the source of $O_2$ and tests the validity of the models and their implications. *Non-detection*: Reconciles the upper limits provided by the remote observations with the in situ detections; Puts clear constraints on the chemical processes in the atmosphere. |
| $H_2S$ | Important component of the sulfur cycle and (in addition to $H_2O$) an important source of hydrogen (a limiting nutrient) for putative aerial biosphere. | Measure altitude-dependent abundance profile of $H_2S$ to 1 ppb precision especially in the clouds and below. Measurements done at several latitudes, by several probes would distinguish between localized vs global distribution of $H_2S$ and inform sources and sinks. | *Detection*: The abundance vs altitude profile constraints the source of $H_2S$ and tests the validity of the models and their implications on the planetary sulfur cycle and overall reservoir of the H-containing species. *Non-detection*: Reconciles the upper limits provided by the remote observations with the tentative in situ detections; Puts clear constraints on the chemical processes in the atmosphere. |
| $NO_x$ | Important components of the planetary nitrogen cycle. | Measure altitude-dependent abundance profile of gaseous $NO_x$ to 1 ppb precision combined with the search for $NO_x$ salt ions within the cloud particles. | *Detection*: The abundance vs altitude profile constraints the source of $NO_x$ and tests the validity of the models and their implications. *Non-detection*: Puts clear constraints on the chemical processes in the cloud droplets and the atmosphere, including on the presence and intensity of lightning strikes. |
| $PH_3$ | Indicator of an unknown chemical processes in the atmosphere and an important member of the planetary phosphorus cycle. | Measure altitude-dependent abundance profile to sub-ppb precision combined with a day and night measurements to inform chemistry sources and sinks. Measurements done at several latitudes, by several probes would distinguish between localized vs global distribution of $PH_3$ and further inform sources and sinks. | *Detection*: The abundance vs altitude profile constraints the source of $PH_3$ and tests the validity of the models and their implications. *Non-detection*: Reconciles the remote and in situ observations with the upper limits; Puts clear constraints on the chemical processes in the atmosphere, including the availability of volatile P species. |
| HCN | Indicator of unknown chemical processes in the clouds. Important precursor for prebiotic chemistry and planetary nitrogen cycle. | Measure altitude-dependent abundance profile of gaseous HCN to 1 ppb precision. | *Detection*: The abundance vs altitude profile constrains the source of HCN and tests the validity of the models and their implications on the planetary nitrogen cycle.<br>*Non-detection*: Reconciles the observational upper limits with the tentative in situ detections. |
| $CH_4$ | Potential sign of life or a result of an unknown abiotic chemical processes on the planet. | Measure altitude-dependent abundance profile, from the top of the clouds down to the surface, of gaseous $CH_4$ to 1 ppb precision. The dedicated instrumentation should be designed to specifically avoid any potential contamination with hydrocarbons brought from Earth or evolved from the instrument itself. | *Detection*: The abundance vs altitude profile constrains the source of $CH_4$ and tests the validity of the models and their implications; Provides a potential source for organic chemistry in the clouds.<br>*Non-detection*: Reconciles the upper limits provided by the remote and in-situ observations with the tentative detection by Pioneer Venus LNMS. |
| Organic molecules | Shows that cloud particles are not chemically simple environment and could contain complex organic molecules that could be precursors to life or even be signs of life itself. | The in-situ search for organic molecules within cloud particles both through detection of fluorescence at multiple wavelengths (e.g. (Baumgardner et al., 2022)), as well as direct identification of organic species to 1 fmol precision from collected cloud particles (in situ (e.g. (Ligterink et al., 2022)) or atmospheric sample return (Seager et al., 2022b)). | *Complex and diverse organics identified*: Potential for life in the cloud particles increases with the diversity and complexity of detected organics.<br>*Only simple and uniform organics identified*: Abiotic processes are most likely responsible for organics formation.<br>*No organics identified*: The prospects of the clouds of Venus as a habitable environment diminish as we assume that all life, no matter its chemical makeup, requires organic chemistry. |
| non-volatiles (e.g. P and Fe) | Cloud particles could contain dissolved metal ions (e.g., Fe) and other ions of non-volatile elements (e.g., P) suggesting that the clouds are not homogenous. Presence of metals could be indicative of efficient interactions between the surface and the clouds. | Qualitative and quantitative elemental analysis and characterization (to 1 ppb precision) of the collected cloud particle material (in situ (e.g. (Ligterink et al., 2022)) or atmospheric sample return (Seager et al., 2022b)). Study interactions between the surface and the atmosphere that might support reservoirs of metals in the clouds. | *Metal ions detected*: The composition of the cloud particles is chemically complex; Suggests efficient exchange of material between the surface (the presumed source of the non-volatile elements) and the clouds.<br>*No metal ions detected*: The material exchange between the surface (the presumed source of the non-volatile elements) and the clouds is not efficient limiting the habitability of the clouds. |
| mode 3 particles | Clouds are not homogenous and are composed of a mixture of particles which may contain different chemistries, including liquid concentrated sulfuric acid and/or solid salt particles. Acidity of cloud particles could be variable and may reach habitable levels. | Chemical analysis of the collected cloud particle material (in situ (e.g. (Ligterink et al., 2022)) or atmospheric sample return (Seager et al., 2022b)). In-situ analysis of particle shape and size distribution (Baumgardner et al., 2022), including direct imaging of particle shapes, direct in-situ determination of the acidity of single cloud particles covering the acidity range from diluted to concentrated sulfuric acid (Kaasik et al., 2022). | *Solid Mode 3 particles composed of salts detected*: Confirms the existence of the Mode 3 particles; the salt composition puts clear constraints on the chemical processes in the cloud droplets and the atmosphere; confirms that the clouds are not uniformly made of liquid concentrated sulfuric acid particles.<br>*No solid particles detected*: Supports the model that the clouds of Venus are made of liquid droplets of concentrated sulfuric acid. |
| | | | *Detection of variable acidity of cloud particles*: The altitude profile of cloud acidity tests the validity of atmospheric and cloud models and model implications for the habitability of the clouds.<br>*Acidity of cloud particles is uniform and consistent with concentrated sulfuric acid*: Puts clear constraints on the chemical processes in the atmosphere; confirms, for the first time by direct measurement, that the clouds are uniformly made of concentrated sulfuric acid particles. |
| unknown absorber | The chemistry of the substance, or substances, absorbing in Venus' clouds is unknown. The unknown absorber could be a sign of biological activity in the clouds. | Global monitoring of the dynamics of the unknown absorber including its spatial and temporal variability (Garvin et al., 2022). Measure the in-situ altitude-dependent UV absorption profile of the clouds. Laboratory and theoretical studies on the absorber candidates, including sulfur species (Francés-Monerris et al., 2022) and organic materials (Spacek, 2021). | *Spatial and temporal variability and quasi-seasonal changes confirmed*: Provide new spectral clues to the nature of the unknown absorption in the upper clouds.<br>*The unknown absorber shows uniform/variable abundance throughout the clouds*: The altitude profile of UV absorption tests the validity of atmospheric and cloud models, including hypotheses on the chemical identity of the unknown absorber. |

**Acknowledgements**

The authors would like to thank Breakthrough Initiatives for the partial funding to conduct this study. We thank the members of the VLF Collaboration (https://venuscloudlife.com/). We also thank Sidney Becker from TU Dortmund University for valuable discussions on the hydroxylamine chemistry. We would like to extend special thanks to the two anonymous reviewers whose comments and suggestions substantially improved our paper.

**Author contributions**

Conceptualization: J.J.P., W.B., S.S., S.R., P.B.R.; methodology: J.J.P., W.B.; analysis W.B., J.J.P.; writing—original draft preparation, J.J.P., S.S. D.H.G.; writing—review and editing: J.J.P., W.B., S.S., S.R., P.B.R., D.H.G., W.P.B., R.A., R.M., C.E.C. All authors have read and agreed to the published version of the manuscript.

**Conflict of interest**

The authors declare no conflict of interest.

**Funding**

This research was partially funded by Breakthrough Initiatives, the Change Happens Foundation, and the Massachusetts Institute of Technology.

**Supplementary Information**

**Table S1.** Free energy per mole for $H_2SO_4/NH_3$ - neutralizing reactions under Venus cloud conditions. Table values revised and updated from (Bains et al., 2021a).

| Reaction | Free energy of reaction (kJ/mol) | Free energy required per mole of $H_2SO_4$ neutralized (kJ/mol) | Water consumed per $H_2SO_4$ neutralized |
|---|---|---|---|
| 1a $4N_{2(aq)} + 9H_2O_{(l)} + 2H_2SO_{4(l)} \rightarrow 3NH_4^+NO_3^-_{(aq)} + 2NH_4^+HSO_4^-_{(aq)}$ | 1420 – 1473 | 710 – 736 | 4.5 |
| 2a $4N_{2(aq)} + 6H_2O_{(l)} + 2H_2SO_{4(l)} \rightarrow 2NH_4^+HSO_4^- + 3H_2O_{2(aq)}$ | 826 – 852 | 413 – 426 | 3 |
| 3a $2N_{2(aq)} + 6H_2O_{(l)} + 4H_2SO_{4(l)} \rightarrow 4NH_4^+HSO_4^- + 3O_{2(aq)}$ | **701 – 752** | **175 – 188** | **1.5** |
| 4a $4N_{2(aq)} + 12H_2O_{(l)} + 3HCl_{(aq)} + 4H_2SO_{4(l)} \rightarrow 5NH_4^+HSO_4^-_{(aq)} + 3NH_4^+ClO_4^-_{(aq)}$ | 1045 – 1141 | 209 – 228 | 3 |

**Table S2.** Free energy per mole for $H_2SO_4/NH_2OH$ - neutralizing reactions under Venus cloud conditions.

| Reaction | Free energy of reaction (kJ/mol) | Free energy required per mole of $H_2SO_4$ neutralized (kJ/mol) | Water consumed per $H_2SO_4$ neutralized |
|---|---|---|---|
| 1b $4N_{2(aq)} + 9H_2O_{(l)} + 2H_2SO_{4(l)} \rightarrow 3NOH_4^+NO_3^-_{(aq)} + 2NOH_4^+HSO_4^-_{(aq)}$ | 1161 – 1257 | 290 – 314 | 4.5 |
| 2b $4N_{2(aq)} + 6H_2O_{(l)} + 2H_2SO_{4(l)} \rightarrow 2NOH_4^+HSO_4^- + 3H_2O_{2(aq)}$ | 599 – 582 | 280 – 291 | 3 |
| 3b $2N_{2(aq)} + 6H_2O_{(l)} + 4H_2SO_{4(l)} \rightarrow 4NOH_4^+HSO_4^- + 3O_{2(aq)}$ | **569 – 617** | **142 – 154** | **1.5** |
| 4b $4N_{2(aq)} + 12H_2O_{(l)} + 3HCl_{(aq)} + 4H_2SO_{4(l)} \rightarrow 5NH_4^+HSO_4^-_{(aq)} + 3NH_4^+ClO_4^-_{(aq)}$ | 724 – 888 | 145 – 178 | 3 |

Results shown in Table S1 and Table S2 are derived for the same conditions as in (Bains et al., 2021a). Free energy of formation of acid solvated $NH_2OH$ derived from gas phase free energy (Bains et al., 2022b), predicted Henry's Constant, and the pKa of hydroxylamine from (Card et al., 2017; Krishna et al., 2003). Note that hydroxylamine nitrate readily decomposes above about 80 °C, in an autocatalytic process (Rafeev and Rubtsov, 1993), and in concentrated or pure form is unstable and liable to explosion (Wei et al., 2006), so it is unlikely to be formed in significant amounts. Hydrolylamine itself is directly dissociated by UV light (Thisuwan et al., 2020). However in the gas phase hydroxylamine absorbs at wavelengths <220 nm (Betts and Back, 1965), and so is likely to be shielded from photolysis in the middle and lower clouds, if it exists there.

**Evidence for a Nitrogen Cycle in the Clouds of Venus**

The recent re-analysis of the Pioneer Venus LNMS data shows evidence of nitrogen chemicals at different oxidation states (from -3 to +5): $NH_3$ (-3), $HCN$ (-3), $N_2$ (0), $NO_2^-$ (+3), $NO_3^-$ (+5).

The potential presence of nitrogen chemicals at different oxidation states implies the existence of an active nitrogen cycle in the clouds of Venus (Figure S1). Such nitrogen compounds could be key electron donors for anoxygenic photosynthesis (nitrite) or a critical redox pair (nitrate and nitrite) for a postulated hypothetical iron-sulfur cycle in Venus' clouds (Limaye et al., 2018). Nitrogen species identified in the re-analyzed data from Pioneer Venus (Mogul et al., 2021b) are also major constituents of the biological nitrogen cycle on Earth (nitrate, nitrite, ammonia, and $N_2$) (Galloway, 2003). The potential identification of $NH_3$, and other N-species from the terrestrial nitrogen cycle is therefore consistent with potential biological activity in the clouds of Venus. We note however that just like on Earth any hypothetical biological production of $NH_3$ (e.g. through fixation of atmospheric $N_2$) would be an energy intensive process, and in the Venusian cloud conditions probably reliant on sunlight (Bains et al., 2021a).

Confirmation of the presence of the multitude of nitrogen species at different oxidation states in the clouds of Venus could establish the existence of an active nitrogen cycle on Venus. The existence of such a cycle will be an important insight into the cloud chemistry, either biotic or abiotic.

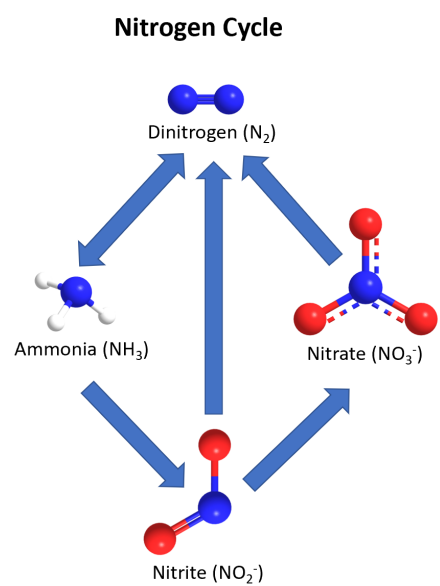

**Figure S1.** The potential nitrogen cycle in the Venusian clouds based on nitrogen species (nitrate, nitrite, ammonia, and $N_2$) tentatively identified in the reanalyzed LNMS data from the Pioneer Venus probe (Mogul et al., 2021b). The identified nitrogen species are also major constituents of the biological nitrogen cycle on Earth.

**References**

Agrawal R, Buchanan WP, Arora A, et al. Mission Architecture to Characterize Habitability of Venus Cloud Layers via an Aerial Platform. Aerospace 2022;9(7):359; doi: 10.3390/aerospace9070359.

Akins AB, Lincowski AP, Meadows VS, et al. Complications in the ALMA Detection of Phosphine at Venus. Astrophys J Lett 2021;907(2):L27.

Albright LF, Houle L, Sumutka AM, et al. Alkylation of Isobutane with Butenes: Effect of Sulfuric Acid Compositions. Ind Eng Chem Process Des Dev 1972;11(3):446–450.

Andreichikov BM. Chemical Composition and Structure of the Clouds of Venus Inferred from the Results of X-Ray Fluorescent Analysis on Descent Probes VEGA 1 and 2. Kosm Issled 1987a;25:721–736.

Andreichikov BM. Chemical Composition and Structure of Venus Clouds from Results of X-Ray Radiometric Experiments Made with the Vega 1 and Vega 2 Automatic Interplanetary Stations. Kosm Issled 1987b;25:737–743.

Baines KH and Delitsky ML. Aqueous Chemistry in the Clouds of Venus: A Possible Source for the UV Absorber. In: AAS/Division for Planetary Sciences Meeting Abstracts# 45 2013; pp. 108–118.

Bains W, Pasek MA, Ranjan S, et al. Large Uncertainties in the Thermodynamics of Phosphorus (III) Oxide (P4O6) Have Significant Implications for Phosphorus Species in Planetary Atmospheres. ACS Earth Sp Chem 2023a;7(6):1219–1226; doi: 10.1021/acsearthspacechem.3c00016.

Bains W, Petkowski JJ, Rimmer PB, et al. Production of Ammonia Makes Venusian Clouds Habitable and Explains Observed Cloud-Level Chemical Anomalies. Proc Natl Acad Sci 2021a;118(52).

Bains W, Petkowski JJ and Seager S. A Data Resource for Sulfuric Acid Reactivity of Organic Chemicals. Data 2021b;6(3):24.

Bains W, Petkowski JJ, Seager S, et al. Phosphine on Venus Cannot Be Explained by Conventional Processes. Astrobiology 2021c;21(10):1277–1304.

Bains W, Petkowski JJ, Seager S, et al. Venusian Phosphine: A 'Wow!' Signal in Chemistry? Phosphorus Sulfur Silicon Relat Elem 2022a;197(5–6):438–443; doi: 10.1080/10426507.2021.1998051.

Bains W, Petkowski JJ and Seager S. Venus' Atmospheric Chemistry and Cloud Characteristics Are Compatible with Venusian Life. Astrobiology 2023b;23(10); doi: 10.1089/ast.2022.0113.

Bains W, Petkowski JJ, Zhan Z, et al. Evaluating Alternatives to Water as Solvents for Life: The Example of Sulfuric Acid. Life 2021d;11(5):400; doi: 10.3390/life11050400.

Bains W, Petkowski JJ, Zhan Z, et al. A Data Resource for Prediction of Gas-Phase Thermodynamic Properties of Small Molecules. Data 2022b;7(3):33.

Bains W, Shorttle O, Ranjan S, et al. Constraints on the Production of Phosphine by Venusian Volcanoes. Universe 2022c;8(1):54.

Bains W, Shorttle O, Ranjan S, et al. Only Extraordinary Volcanism Can Explain the Presence of Parts per Billion Phosphine on Venus. Proc Natl Acad Sci 2022d;119(7):e2121702119.

Baumgardner D, Fisher T, Newton R, et al. Deducing the Composition of Venus Cloud Particles with the Autofluorescence Nephelometer (AFN). Aerospace 2022;9(9):492.

Belyaev DA, Montmessin F, Bertaux J-L, et al. Vertical Profiling of SO2 and SO above Venus' Clouds by SPICAV/SOIR Solar Occultations. Icarus 2012;217(2):740–751.

Benner SA and Spacek J. The Limits to Organic Life in the Solar System: From Cold Titan to Hot Venus. LPI Contrib 2021;2629:4003.

De Bergh C, Bezard B, Crisp D, et al. Water in the Deep Atmosphere of Venus from High-Resolution Spectra of the Night Side. Adv Sp Res 1995;15(4):79–88.

Bertaux J, Widemann T, Hauchecorne A, et al. VEGA 1 and VEGA 2 Entry Probes: An Investigation of Local UV Absorption (220–400 Nm) in the Atmosphere of Venus (SO2 Aerosols, Cloud Structure). J Geophys Res Planets

1996;101(E5):12709–12745.

Betts J and Back RA. The Photolysis and the Mercury-Photosensitized Decomposition of Hydroxylamine Vapor. Can J Chem 1965;43(10):2678–2684.

Bézard B, Tsang CCC, Carlson RW, et al. Water Vapor Abundance near the Surface of Venus from Venus Express/VIRTIS Observations. J Geophys Res Planets 2009;114(E5); doi: https://doi.org/10.1029/2008JE003251.

Bierson CJ and Zhang X. Chemical Cycling in the Venusian Atmosphere: A Full Photochemical Model from the Surface to 110 Km. J Geophys Res Planets 2019;e2019JE006.

Blackie D, Blackwell-Whitehead R, Stark G, et al. High-resolution Photoabsorption Cross-section Measurements of $SO_2$ at 198 K from 213 to 325 Nm. J Geophys Res Planets 2011;116(E3).

Bornstein P and Balian G. Cleavage at Asn-Gly Bonds with Hydroxylamine. In: Methods in Enzymology Elsevier; 1977; pp. 132–145.

Buchanan WP, de Jong M, Agrawal R, et al. Aerial Platform Design Options for a Life-Finding Mission at Venus. Aerospace 2022;9(7):363; doi: 10.3390/aerospace9070363.

Busenberg E and Plummer LN. Dating Young Groundwater with Sulfur Hexafluoride: Natural and Anthropogenic Sources of Sulfur Hexafluoride. Water Resour Res 2000;36(10):3011–3030.

Cabrol NA, Wettergreen D, Warren-Rhodes K, et al. Life in the Atacama: Searching for Life with Rovers (Science Overview). J Geophys Res Biogeosciences 2007;112(G4).

Card ML, Gomez-Alvarez V, Lee W-H, et al. History of EPI Suite™ and Future Perspectives on Chemical Property Estimation in US Toxic Substances Control Act New Chemical Risk Assessments. Environ Sci Process Impacts 2017;19(3):203–212.

Cheng M-D. Atmospheric Chemistry of Hydrogen Fluoride. J Atmos Chem 2018;75(1):1–16.

Cleland CE. Moving beyond Definitions in the Search for Extraterrestrial Life. Astrobiology 2019a;19(6):722–729.

Cleland CE. The Quest for a Universal Theory of Life: Searching for Life as We Don't Know It. Cambridge University Press; 2019b.

Cleland CE and Rimmer PB. Ammonia and Phosphine in the Clouds of Venus as Potentially Biological Anomalies. Aerospace 2022;9(12):752.

Clements DL. Venus, Phosphine and the Possibility of Life. Contemp Phys 2022;63(3):180–99; doi: 10.1080/00107514.2023.2184932.

Cockell CS. Life on Venus. Planet Space Sci 1999;47(12):1487–1501.

Cordiner MA, Villanueva GL, Wiesemeyer H, et al. Phosphine in the Venusian Atmosphere: A Strict Upper Limit from SOFIA GREAT Observations. Geophys Res Lett 2022;e2022GL101055.

Dai L, Zhang X, Shao WD, et al. A Simple Condensation Model for the H2SO4-H2O Gas-cloud System on Venus. J Geophys Res Planets 2022;127:e2021JE007060.

Dartnell LR, Nordheim TA, Patel MR, et al. Constraints on a Potential Aerial Biosphere on Venus: I. Cosmic Rays. Icarus 2015;257:396–405.

Donahue TM and Hodges Jr RR. Past and Present Water Budget of Venus. J Geophys Res Planets 1992;97(E4):6083–6091.

Donahue TM and Hodges Jr RR. Venus Methane and Water. Geophys Res Lett 1993;20(7):591–594.

Drossart P, Bézard B, Encrenaz T, et al. Search for Spatial Variations of the H2O Abundance in the Lower Atmosphere of Venus from NIMS-Galileo. Planet Space Sci 1993;41(7):495–504.

Ekonomov AP, Moroz VI, Moshkin BE, et al. Scattered UV Solar Radiation within the Clouds of Venus. Nature 1984;307(5949):345–347.

Ekonomov AP, Moshkin BE, Moroz VI, et al. UV Photometry at the Venera 13 and 14 Landing Probes. Cosm Res 1983;21(2):194–206.

Encrenaz T, Greathouse TK, Marcq E, et al. A Stringent Upper Limit of the PH3 Abundance at the Cloud Top of Venus. Astron Astrophys 2020;643:L5.

Esposito L, Knollenberg R, Marov M, et al. 16. the Clouds and Hazes of Venus. Venus 1983;484.

Esposito LW, Bertaux JL, Krasnopolsky VA, et al. Chemistry of Lower Atmosphere and Clouds. Venus II 1997;415–458.

Fimmel RO. Pioneer Venus. NASA Scientific and Technical Information 329 Branch: Washington D.C.; 1983.

Florensky CP, Volkov VP and Nikolaeva O V. A Geochemical Model of the Venus Troposphere. Icarus 1978;33(3):537–553.

Francés-Monerris A, Carmona-García J, Trabelsi T, et al. Photochemical and Thermochemical Pathways to S2 and Polysulfur Formation in the Atmosphere of Venus. Nat Commun 2022;13(1):1–8.

Frandsen BN, Farahani S, Vogt E, et al. Spectroscopy of OSSO and Other Sulfur Compounds Thought to Be Present in the Venus Atmosphere. J Phys Chem A 2020;124(35):7047–7059.

Frandsen BN, Wennberg PO and Kjaergaard HG. Identification of OSSO as a Near-UV Absorber in the Venusian Atmosphere. Geophys Res Lett 2016;43(21):11–146.

Freeman A, Smrekar SE, Hensley S, et al. Veritas: A Discovery-Class Venus Surface Geology and Geophysics Mission. 2016.

French R, Mandy C, Hunter R, et al. Rocket Lab Mission to Venus. Aerospace 2022;9(8):445; doi: 10.3390/aerospace9080445.

Galloway JN. The Global Nitrogen Cycle. Treatise on Geochemistry 2003;8–9:557–583; doi: 10.1016/B0-08-043751-6/08160-3.

Garvin JB, Getty SA, Arney GN, et al. Revealing the Mysteries of Venus: The DAVINCI Mission. Planet Sci J 2022;3(5):117; doi: 10.3847/psj/ac63c2.

Gelman BG, Drozdov Y V, Melnikov V V, et al. Reaction Gas Chromatography of Venus Cloud Aerosols. Sov Astron Lett 1986;12:42.

Gelman BG, Zolotukhin VG, Lamonov NI, et al. Gas-Chromatographical Analysis of the Chemical Composition of the Venus Atmosphere Made with the Venera 12 Probe. Pisma v Astron Zhurnal 1979a;5:217–221.

Gelman BG, Zolotukhin VG, Lamonov NI, et al. Venera 12 Analysis of Venus Atmospheric Composition by Gas Chromatography. Sov Astron Lett 1979b;5:116–118.

Gillmann C, Way MJ, Avice G, et al. The Long-Term Evolution of the Atmosphere of Venus: Processes and Feedback Mechanisms. Space Sci Rev 2022;218(7):56; doi: 10.1007/s11214-022-00924-0.

Greaves JS, Bains W, Petkowski JJ, et al. Addendum: Phosphine Gas in the Cloud Deck of Venus. Nat Astron 2021a;5(7):726–728.

Greaves JS, Petkowski JJ, Richards A, et al. Recovery of Phosphine in Venus' Atmosphere from SOFIA Observations. arXiv Prepr arXiv221109852 2022a.

Greaves JS, Richards AMS, Bains W, et al. Phosphine Gas in the Cloud Decks of Venus. Nat Astron 2021b;5(7):655–664.

Greaves JS, Richards AMS, Bains W, et al. Reply to: No Evidence of Phosphine in the Atmosphere of Venus from Independent Analyses. Nat Astron 2021c;5(7):636–639.

Greaves JS, Rimmer PB, Richards A, et al. Low Levels of Sulphur Dioxide Contamination of Venusian Phosphine Spectra. Mon Not R Astron Soc 2022b;514(2):2994–3001; doi: 10.1093/mnras/stac1438.

Greenwood NN and Earnshaw A. Chemistry of the Elements. Elsevier: Amsterdam; 2012.

Grinspoon DH. Venus Revealed: A New Look below the Clouds of Our Mysterious Twin Planet. 1997.

Grinspoon DH and Bullock MA. Astrobiology and Venus Exploration. Geophys Monogr Geophys Union 2007;176:191.

Hansen JE and Hovenier JW. Interpretation of the Polarization of Venus. J Atmos Sci 1974;31(4):1137–1160.

Hapke B and Graham F. Spectral Properties of Condensed Phases of Disulfur Monoxide, Polysulfur Oxide, and Irradiated Sulfur. Icarus 1989;79(1):47–55.

Hapke B and Nelson R. Evidence for an Elemental Sulfur Component of the Clouds from Venus Spectrophotometry. J Atmos Sci 1975;32(6):1212–1218.

Harnisch J and Eisenhauer A. Natural CF4 and SF6 on Earth. Geophys Res Lett 1998;25(13):2401–2404.

Hartley KK, Wolff AR and Travis LD. Croconic Acid: An Absorber in the Venus Clouds? Icarus 1989;77(2):382–390.

Hoffman JH, Hodges RR, Donahue TM, et al. Composition of the Venus Lower Atmosphere from the Pioneer Venus Mass Spectrometer. J Geophys Res Sp Phys 1980a;85(A13):7882–7890.

Hoffman JH, Oyama VI and Von Zahn U. Measurements of the Venus Lower Atmosphere Composition: A Comparison of Results. J Geophys Res Sp Phys 1980b;85(A13):7871–7881.

Huang Q, Zhao G, Zhang S, et al. Improved Catalytic Lifetime of H2SO4 for Isobutane Alkylation with Trace Amount of Ionic Liquids Buffer. Ind Eng Chem Res 2015;54(5):1464–1469.

Huntzicker JJ, Cary RA and Ling C-S. Neutralization of Sulfuric Acid Aerosol by Ammonia. Environ Sci Technol 1980;14(7):819–824.

Ignatiev NI, Moroz VI, Moshkin BE, et al. Water Vapour in the Lower Atmosphere of Venus: A New Analysis of Optical Spectra Measured by Entry Probes. Planet Space Sci 1997;45(4):427–438.

Istomin VG, Grechnev K V and Kochnev VA. Mass Spectrometry Measurements of the Lower Atmosphere of Venus. 1979a.

Istomin VG, Grechnev K V and Kochnev VA. Venera 11 and 12 Mass Spectrometry of the Lower Venus Atmosphere. Sov Astron Lett 1979b;5:113.

Istomin VG, Grechnev K V and Kochnev VA. Mass-Spectrometry of the Atmosphere by Venera 13 and Venera 14. Cosm Res 1983;21(3):329–338.

Istomin VG, Grechnev K V and Kotchnev VA. Mass Spectrometer Measurements of the Composition of the Lower Atmosphere of Venus. In: COSPAR Colloquia Series Elsevier; 1980; pp. 215–218.

Izenberg NR, Gentry DM, Smith DJ, et al. The Venus Life Equation. Astrobiology 2021;21(10):1305–1315.

James EP, Toon OB and Schubert G. A Numerical Microphysical Model of the Condensational Venus Cloud. Icarus 1997;129(1):147–171.

Johnson NM and de Oliveira MRR. Venus Atmospheric Composition in Situ Data: A Compilation. Earth Sp Sci 2019;6(7):1299–1318.

Jordan S, Shorttle O and Rimmer PB. Proposed Energy-Metabolisms Cannot Explain the Atmospheric Chemistry of Venus. Nat Commun 2022;13(1):3274; doi: 10.1038/s41467-022-30804-8.

Kaasik L, Rahu I, Roper EM, et al. Sensor for Determining Single Droplet Acidities in the Venusian Atmosphere. Aerospace 2022;9(10):560; doi: 10.3390/aerospace9100560.

Keller-Rudek H, Moortgat GK, Sander R, et al. The MPI-Mainz UV/VIS Spectral Atlas of Gaseous Molecules of Atmospheric Interest. Earth Syst Sci Data 2013;5(2):365–373; doi: 10.5194/essd-5-365-2013.

Knollenberg R, Travis L, Tomasko M, et al. The Clouds of Venus: A Synthesis Report. J Geophys Res Sp Phys 1980;85(A13):8059–8081.

Knollenberg RG. Clouds and Hazes. Nature 1982;296(5852):18.

Knollenberg RG. A Reexamination of the Evidence for Large, Solid Particles in the Clouds of Venus. Icarus 1984;57(2):161–183.

Knollenberg RG and Hunten DM. Clouds of Venus: Particle Size Distribution Measurements. Science (80- ) 1979;203(4382):792–795.

Knollenberg RG and Hunten DM. The Microphysics of the Clouds of Venus: Results of the Pioneer Venus Particle Size Spectrometer Experiment. J Geophys Res Sp Phys 1980;85(A13):8039–8058.

Kotsyurbenko OR, Cordova JA, Belov AA, et al. Exobiology of the Venusian Clouds: New Insights into Habitability through Terrestrial Models and Methods of Detection. Astrobiology 2021.

Krasnopolsky V. Observation of DCl and Upper Limit to NH3 on Venus. Icarus 2012a;219(1):244–249.

Krasnopolsky VA. Chemical Composition of Venus Clouds. Planet Space Sci 1985;33(1):109–117.

Krasnopolsky VA. Photochemistry of the Atmospheres of Mars and Venus. 1st ed. (Zahn U von. ed). Springer-Verlag: New York; 1986.

Krasnopolsky VA. Vega Mission Results and Chemical Composition of Venusian Clouds. Icarus 1989;80(1):202–210; doi: https://doi.org/10.1016/0019-1035(89)90168-1.

Krasnopolsky VA. Chemical Composition of Venus Atmosphere and Clouds: Some Unsolved Problems. Planet Space Sci 2006;54(13–14):1352–1359.

Krasnopolsky VA. Chemical Kinetic Model for the Lower Atmosphere of Venus. Icarus 2007;191(1):25–37.

Krasnopolsky VA. High-Resolution Spectroscopy of Venus: Detection of OCS, Upper Limit to H2S, and Latitudinal Variations of CO and HF in the Upper Cloud Layer. Icarus 2008;197(2):377–385.

Krasnopolsky VA. A Photochemical Model for the Venus Atmosphere at 47–112 Km. Icarus 2012b;218(1):230–246.

Krasnopolsky VA. S3 and S4 Abundances and Improved Chemical Kinetic Model for the Lower Atmosphere of Venus. Icarus 2013;225(1):570–580.

Krasnopolsky VA. Vertical Profiles of H2O, H2SO4, and Sulfuric Acid Concentration at 45–75 Km on Venus. Icarus 2015;252:327–333.

Krasnopolsky VA. Sulfur Aerosol in the Clouds of Venus. Icarus 2016;274:33–36.

Krasnopolsky VA. On the Iron Chloride Aerosol in the Clouds of Venus. Icarus 2017;286:134–137.

Krasnopolsky VA. Disulfur Dioxide and Its Near-UV Absorption in the Photochemical Model of Venus Atmosphere. Icarus 2018;299:294–299.

Krasnopolsky VA and Belyaev DA. Search for HBr and Bromine Photochemistry on Venus. Icarus 2017;293:114–118.

Krishna K, Wang Y, Saraf SR, et al. Hydroxylamine Production: Will a QRA Help You Decide? Reliab Eng Syst Saf 2003;81(2):215–224.

Krissansen-Totton J, Bergsman DS and Catling DC. On Detecting Biospheres from Chemical Thermodynamic Disequilibrium in Planetary Atmospheres. Astrobiology 2016;16(1):39–67.

Lee YJ, Jessup K-L, Perez-Hoyos S, et al. Long-Term Variations of Venus's 365 Nm Albedo Observed by Venus Express, Akatsuki, MESSENGER, and the Hubble Space Telescope. Astron J 2019;158(3):126.

Ligterink NFW, Kipfer KA, Gruchola S, et al. The ORIGIN Space Instrument for Detecting Biosignatures and Habitability Indicators on a Venus Life Finder Mission. Aerospace 2022;9(6):312.

Limaye S, Bullock MA, Baines KH, et al. Venus, an Astrobiology Target. Bull AAS 2021a;18(53); doi: 10.3847/25c2cfeb.32998123.

Limaye SS and Garvin JB. Exploring Venus: Next Generation Missions beyond Those Currently Planned. Front Astron Sp Sci 2023;10:1188096.

Limaye SS, Mogul R, Baines KH, et al. Venus, an Astrobiology Target. Astrobiology 2021b;21(10):1163–1185.

Limaye SS, Mogul R, Smith DJ, et al. Venus' Spectral Signatures and the Potential for Life in the Clouds. Astrobiology 2018;18(9):1181–1198.

Lincowski AP, Meadows VS, Crisp D, et al. Claimed Detection of PH3 in the Clouds of Venus Is Consistent with Mesospheric SO2. Astrophys J Lett 2021;908:L44-52.

Mahieux A, Viscardy S, Jessup KL, et al. H2CO, O3, NH3, HCN, N2O, NO2, NO, and HO2 Upper Limits of Detection in the Venus Lower-Mesosphere Using SOIR on Board Venus Express. Icarus 2023;115862.

Marcq E, Mills FP, Parkinson CD, et al. Composition and Chemistry of the Neutral Atmosphere of Venus. Space Sci Rev 2018;214(1):10.

Marov MY and Grinspoon DH. The Planet Venus. Yale University Press; 1998.

Mills FP. I. Observations and Photochemical Modeling of the Venus Middle Atmosphere. II. Thermal Infrared

Spectroscopy of Europa and Callisto. California Institute of Technology; 1998.

Mills FP. A Spectroscopic Search for Molecular Oxygen in the Venus Middle Atmosphere. J Geophys Res Planets 1999;104(E12):30757–30763.

Mills FP, Esposito LW and Yung YL. Atmospheric Composition, Chemistry, and Clouds. 2007.

Milojevic T, Treiman AH and Limaye S. Phosphorus in the Clouds of Venus: Potential for Bioavailability. Astrobiology 2021.

Miron S and Lee RJ. Molecular Structure of Conjunct Polymers. J Chem Eng Data 1963;8(1):150–160.

Mogul R, Limaye SS, Lee YJ, et al. Potential for Phototrophy in Venus' Clouds. Astrobiology 2021a;21(10):1237–1249; doi: 10.1089/ast.2021.0032.

Mogul R, Limaye SS, Way MJ, et al. Venus' Mass Spectra Show Signs of Disequilibria in the Middle Clouds. Geophys Res Lett 2021b;e2020GL091327.

Mogul R, Limaye SS and Way MJ. The CO2 Profile and Analytical Model for the Pioneer Venus Large Probe Neutral Mass Spectrometer. Icarus 2022;115374; doi: https://doi.org/10.1016/j.icarus.2022.115374.

Morowitz H and Sagan C. Life in the Clouds of Venus? Nature 1967;215(5107):1259–1260; doi: 10.1038/2151259a0.

Moroz VI. The Atmosphere of Venus. Space Sci Rev 1981;29(1):3–127.

Moshkin BE, Moroz VI, Gnedykh VI, et al. VEGA-1 and VEGA-2 Optical Spectrometry of Venus Atmospheric Aerosols at the 60-30-KM Levels-Preliminary Results. Sov Astron Lett 1986;12:36–39.

Mukhin LM, Gelman BG, Lamonov NI, et al. VENERA-13 and VENERA-14 Gas Chromatography Analysis of the Venus Atmosphere Composition. Sov Astron Lett 1982;8:216–218.

Mukhin LM, Nenarokov DF, Porschnev N V, et al. Preliminary Calibration Results of Vega 1 and 2 SIGMA-3 Gas Chromatograph. Adv Sp Res 1987;7(12):329–335.

Nakamura M, Imamura T, Ishii N, et al. Overview of Venus Orbiter, Akatsuki. Earth, planets Sp 2011;63(5):443–457.

Newton AS. Some Observations on the Mass Spectrum of CO2. J Chem Phys 1952;20(8):1330–1331.

de Oliveira MRR, Gil PJS and Ghail R. A Novel Orbiter Mission Concept for Venus with the EnVision Proposal. Acta Astronaut 2018;148:260–267.

Omran A, Oze C, Jackson B, et al. Phosphine Generation Pathways on Rocky Planets. Astrobiology 2021;21(10):1264–1276; doi: 10.1089/ast.2021.0034.

Oschlisniok J, Häusler B, Pätzold M, et al. Microwave Absorptivity by Sulfuric Acid in the Venus Atmosphere: First Results from the Venus Express Radio Science Experiment VeRa. Icarus 2012;221(2):940–948; doi: https://doi.org/10.1016/j.icarus.2012.09.029.

Oschlisniok J, Häusler B, Pätzold M, et al. Sulfuric Acid Vapor and Sulfur Dioxide in the Atmosphere of Venus as Observed by the Venus Express Radio Science Experiment VeRa. Icarus 2021;362:114405.

Oyama VI, Carle GC, Woeller F, et al. Laboratory Corroboration of the Pioneer Venus Gas Chromatograph Analyses. Science (80- ) 1979a;205(4401):52–54.

Oyama VI, Carle GC, Woeller F, et al. Venus Lower Atmospheric Composition: Analysis by Gas Chromatography. Science (80- ) 1979b;203(4382):802–805.

Oyama VI, Carle GC and Woeller F. Corrections in the Pioneer Venus Sounder Probe Gas Chromatographic Analysis of the Lower Venus Atmosphere. Science (80- ) 1980a;208(4442):399–401.

Oyama VI, Carle GC, Woeller F, et al. Pioneer Venus Gas Chromatography of the Lower Atmosphere of Venus. J Geophys Res Sp Phys 1980b;85(A13):7891–7902.

Oyama VI, Carle GC, Woeller F, et al. Pioneer Venus Sounder Probe Gas Chromatograph. IEEE Trans Geosci Remote Sens 1980c;18:85–93.

Palmer KF and Williams D. Optical Constants of Sulfuric Acid; Application to the Clouds of Venus? Appl Opt 1975;14(1):208–219.

Parro V, de Diego-Castilla G, Moreno-Paz M, et al. A Microbial Oasis in the Hypersaline Atacama Subsurface Discovered by a Life Detector Chip: Implications for the Search for Life on Mars. Astrobiology 2011;11(10):969–996.

Patel MR, Mason JP, Nordheim TA, et al. Constraints on a Potential Aerial Biosphere on Venus: II. Ultraviolet Radiation. Icarus 2021;114796; doi: https://doi.org/10.1016/j.icarus.2021.114796.

Pérez-Hoyos S, Sánchez-Lavega A, García-Muñoz A, et al. Venus Upper Clouds and the UV Absorber from MESSENGER/MASCS Observations. J Geophys Res Planets 2018;123(1):145–162.

Petrianov I V, Andreichikov BM, Korchuganov BN, et al. Iron in the Clouds of Venus. In: Akademiia Nauk SSSR Doklady 1981; pp. 834–836.

Pollack JB, Dalton JB, Grinspoon D, et al. Near-Infrared Light from Venus' Nightside: A Spectroscopic Analysis. Icarus 1993;103(1):1–42.

Pollack JB, Ragent B, Boese R, et al. Nature of the Ultraviolet Absorber in the Venus Clouds: Inferences Based on Pioneer Venus Data. Science (80- ) 1979;205(4401):76–79.

Pollack JB, Toon OB, Whitten RC, et al. Distribution and Source of the UV Absorption in Venus' Atmosphere. J Geophys Res Sp Phys 1980;85(A13):8141–8150.

Porshnev N V, Mukhin LM, Gel'Man BG, et al. Gas-Chromatographic Analysis of Products of Thermal Reactions of Aerosol in the Venus Cloud Layer on Board Vega 1 and Vega 2. Kosm Issled 1987;25:715–720.

Porshnev N V, Mukhin LM, Gelman BG, et al. Gas-Chromatographic Analysis of Products of Thermal Reactions of Aerosol in Venusian Cloud Layer on Vega-1 and Vega-2 Automatic Interplanetary Stations. JPRS Rep Sci Technol USSR Sp 1988;25(5):16.

Queen L, Getty S, Johnson N, et al. DAVINCI In Situ Capability Roundtable. 2022. Available from: https://ssed.gsfc.nasa.gov/davinci/roundtable [Last accessed: 7/28/2022].

Rafeev VA and Rubtsov YI. Kinetics and Mechanism of Thermal Decomposition of Hydroxylammonium Nitrate. Russ Chem Bull 1993;42(11):1811–1815.

Ragent B and Blamont J. Preliminary Results of the Pioneer Venus Nephelometer Experiment. Science (80- ) 1979;203(4382):790–792.

Rimmer PB, Jordan S, Constantinou T, et al. Hydroxide Salts in the Clouds of Venus: Their Effect on the Sulfur Cycle and Cloud Droplet PH. Planet Sci J 2021;2(4):133.

Ross FE. Photographs of Venus. Astrophys J 1928;68:57.

Rustad DS and Gregory NW. Gas-Phase Ultraviolet and Visible Spectra of Sodium Tetrachloroferrate (III) and of Monomeric and Dimeric Iron (III) Chloride. Inorg Chem 1977;16(12):3036–3040.

Sagan C, Thompson WR, Carlson R, et al. A Search for Life on Earth from the Galileo Spacecraft. Nature 1993;365(6448):715–721.

Sandor BJ, Clancy RT, Moriarty-Schieven G, et al. Sulfur Chemistry in the Venus Mesosphere from $SO_2$ and SO Microwave Spectra. Icarus 2010;208(1):49–60.

Schulze-Makuch D, Grinspoon DH, Abbas O, et al. A Sulfur-Based Survival Strategy for Putative Phototrophic Life in the Venusian Atmosphere. Astrobiology 2004;4(1):11–18.

Schulze-Makuch D and Irwin LN. Reassessing the Possibility of Life on Venus: Proposal for an Astrobiology Mission. Astrobiology 2002;2(2):197–202; doi: 10.1089/15311070260192264.

Schulze-Makuch D and Irwin LN. The Prospect of Alien Life in Exotic Forms on Other Worlds. Naturwissenschaften 2006;93(4):155–172.

Seager S, Petkowski JJ, Carr CE, et al. Venus Life Finder Habitability Mission: Motivation, Science Objectives, and Instrumentation. Aerospace 2022a;9(11):733; doi: 10.3390/aerospace9110733.

Seager S, Petkowski JJ, Carr CE, et al. Venus Life Finder Missions Motivation and Summary. Aerospace 2022b;9(7):385; doi: 10.3390/aerospace9070385.

Seager S, Petkowski JJ, Gao P, et al. The Venusian Lower Atmosphere Haze as a Depot for Desiccated Microbial Life: A Proposed Life Cycle for Persistence of the Venusian Aerial Biosphere. Astrobiology 2021;21(10):1206–1223.

Seager S, Petkowski JJ, Huang J, et al. Fully Fluorinated Non-Carbon Compounds NF3 and SF6 as Ideal Technosignature Gases. Sci Rep 2023a;13(1):13576.

Seager S, Petkowski JJ, Seager MD, et al. Stability of Nucleic Acid Bases in Concentrated Sulfuric Acid: Implications for the Habitability of Venus' Clouds. Proc Natl Acad Sci 2023b;120(25):e2220007120; doi: 10.1073/pnas.2220007120.

Shaya E and Caldwell J. Photometry of Venus below 2200 Å from the Orbiting Astronomical Observatory-2. Icarus 1976;27(2):255–264.

Shimizu M. Ultraviolet Absorbers in the Venus Clouds. Astrophys Space Sci 1977;51(2):497–499.

Sill GT. The Composition of the Ultraviolet Dark Markings on Venus. J Atmos Sci 1975;32(6):1201–1204.

Simmonds PG, Rigby M, Manning AJ, et al. The Increasing Atmospheric Burden of the Greenhouse Gas Sulfur Hexafluoride (SF 6). Atmos Chem Phys 2020;20(12):7271–7290.

Snellen IAG, Guzman-Ramirez L, Hogerheijde MR, et al. Re-Analysis of the 267 GHz ALMA Observations of Venus-No Statistically Significant Detection of Phosphine. Astron Astrophys 2020;644:L2.

Spacek J. Organic Carbon Cycle in the Atmosphere of Venus. arXiv Prepr arXiv210802286 2021.

Spacek J and Benner SA. The Organic Carbon Cycle in the Atmosphere of Venus and Evolving Red Oil. LPI Contrib 2021;2629:4052.

Spacek J, Rimmer P, Cady S, et al. Organics Produced in the Clouds of Venus Resemble the Spectrum of the Unknown Absorber. LPI Contrib 2023;2807:8060.

Surkov A, Andreichikov BM and Kalinkina OM. Composition and Structure of the Cloud Layer of Venus. In: Space Research, XIV: Proceedings of Open Meetings of Working Groups of the 16th Plenary Meeting of COSPAR Constance; 1974; pp. 673–678.

Surkov IA. Geochemical Studies of Venus by Venera 9 and 10 Automatic Interplanetary Stations. In: Lunar and Planetary Science Conference Proceedings 1977; pp. 2665–2689.

Surkov IA, Andreichikov BM and Kalinkina OM. Gas-Analysis Equipment for the Automatic Interplanetary Stations Venera-4, 5, 6 and 8. Sp Sci Instrum 1977;3:301–310.

Surkov IA, Ivanova VF, Pudov AN, et al. Water Vapour Content in the Venus Atmosphere from Venera 13 and Venera 14 Data. Kosm Issled 1983;21:231–235.

Surkov IUA, Shcheglov OP, Ryvkin ML, et al. Distribution of Water Vapor in the Middle and Lower Atmosphere of Venus. Kosm Issled 1987;25:678–690.

Surkov YA, Andrejchikov BM and Kalinkina OM. On the Content of Ammonia in the Venus Atmosphere Based on Data Obtained from Venera 8 Automatic Station. In: Akademiia Nauk SSSR Doklady 1973; pp. 296–298.

Surkov YA, Ivanova VF, Pudov AN, et al. VEGA-1 Mass Spectrometry of Venus Cloud Aerosols-Preliminary Results. Sov Astron Lett 1986;12:44.

Surkov YA, Kirnozov FF, Glasov VN, et al. New Data on the Venus Cloud Aerosol (Preliminary Results of Measurements by the Venera 14 Probe). Pis' ma Astron Zh 1982;8:700.

Svedhem H, Titov D, Taylor F, et al. Venus Express Mission. J Geophys Res Planets 2009;114(E5).

Thisuwan J, Promma P and Sagarik K. Photolytic Mechanisms of Hydroxylamine. RSC Adv 2020;10(14):8319–8331.

Thompson MA. The Statistical Reliability of 267-GHz JCMT Observations of Venus: No Significant Evidence for Phosphine Absorption. Mon Not R Astron Soc Lett 2021;501(1):L18–L22.

Titov D V. On the Possibility of Aerosol Formation by the Reaction between $SO_2$ and $NH_3$ in the Venus Atmosphere. Cosm Res 1983;21:401–408.

Titov D V, Ignatiev NI, McGouldrick K, et al. Clouds and Hazes of Venus. Space Sci Rev 2018;214(8):1–61.

Titov D V, Markiewicz WJ, Ignatiev NI, et al. Morphology of the Cloud Tops as Observed by the Venus Express

Monitoring Camera. Icarus 2012;217(2):682–701.

Tomasko MG, Doose LR and Smith PH. Absorption of Sunlight in the Atmosphere of Venus. Science (80- ) 1979;205(4401):80–82.

Tomasko MG, Doose LR, Smith PH, et al. Measurements of the Flux of Sunlight in the Atmosphere of Venus. J Geophys Res Sp Phys 1980;85(A13):8167–8186.

Toon OB, Ragent B, Colburn D, et al. Large, Solid Particles in the Clouds of Venus: Do They Exist? Icarus 1984;57(2):143–160.

Toon OB, Turco RP and Pollack JB. The Ultraviolet Absorber on Venus: Amorphous Sulfur. Icarus 1982;51(2):358–373.

Trauger JT and Lunine JI. Spectroscopy of Molecular Oxygen in the Atmospheres of Venus and Mars. Icarus 1983;55(2):272–281.

Trompet L, Robert S, Mahieux A, et al. Phosphine in Venus' Atmosphere: Detection Attempts and Upper Limits above the Cloud Top Assessed from the SOIR/VEx Spectra. Astron Astrophys 2020;645:L4.

Truong N and Lunine JI. Volcanically Extruded Phosphides as an Abiotic Source of Venusian Phosphine. Proc Natl Acad Sci 2021;118(29).

Vandaele AC, Korablev O, Belyaev D, et al. Sulfur Dioxide in the Venus Atmosphere: I. Vertical Distribution and Variability. Icarus 2017;295:16–33.

Villanueva GL, Cordiner M, Irwin PGJ, et al. No Evidence of Phosphine in the Atmosphere of Venus from Independent Analyses. Nat Astron 2021;5(7):631–635.

Vinogradov AP, Surkov A and Andreichikov BM. Research in the Composition of the Atmosphere of Venus Aboard Automatic Stations" Venera-5" and" Venera-6". In: Soviet Physics Doklady 1970; p. 4.

Vítek P, Jehlička J, Edwards HGM, et al. The Miniaturized Raman System and Detection of Traces of Life in Halite from the Atacama Desert: Some Considerations for the Search for Life Signatures on Mars. Astrobiology 2012;12(12):1095–1099.

Waltham C, Boyle J, Ramey B, et al. Light Scattering and Absorption Caused by Bacterial Activity in Water. Appl Opt 1994;33(31):7536–7540.

Watson AJ, Donahue TM, Stedman DH, et al. Oxides of Nitrogen and the Clouds of Venus. Geophys Res Lett 1979;6(9):743–746.

Way MJ and Del Genio AD. Venusian Habitable Climate Scenarios: Modeling Venus Through Time and Applications to Slowly Rotating Venus-Like Exoplanets. J Geophys Res Planets 2020;125(5):e2019JE006276.

Wei C, Rogers WJ and Mannan MS. Thermal Decomposition Hazard Evaluation of Hydroxylamine Nitrate. J Hazard Mater 2006;130(1–2):163–168.

Winick JR and Stewart AIF. Photochemistry of $SO_2$ in Venus' Upper Cloud Layers. J Geophys Res Sp Phys 1980;85(A13):7849–7860.

Wu Z, Wan H, Xu J, et al. The Near-UV Absorber OSSO and Its Isomers. Chem Commun 2018;54(36):4517–4520; doi: 10.1039/C8CC00999F.

Wunderlich F, Grenfell JL and Rauer H. Uncertainty in Phosphine Photochemistry in the Venus Atmosphere Prevents a Firm Biosignature Attribution. Astron Astrophys 2023;676:A135.

Yamazaki A, Yamada M, Lee YJ, et al. Ultraviolet Imager on Venus Orbiter Akatsuki and Its Initial Results. Earth, Planets Sp 2018;70(1):1–11.

Young AT. Are the Clouds of Venus Sulfuric Acid? Icarus 1973;18(4):564–582; doi: 10.1016/0019-1035(73)90059-6.

Young AT. An Improved Venus Cloud Model. Icarus 1977;32(1):1–26.

Yung YL and Demore WB. Photochemistry of the Stratosphere of Venus: Implications for Atmospheric Evolution. Icarus 1982;51(2):199–247; doi: 10.1016/0019-1035(82)90080-X.

Von Zahn U, Kumar S, Niemann H, et al. 13. Composition of the Venus Atmosphere. Venus 1983;299.

Von Zahn U and Moroz VI. Composition of the Venus Atmosphere below 100 Km Altitude. Adv Sp Res 1985;5(11):173–195.

Zasova L V, Krasnopolsky VA and Moroz VI. Vertical Distribution of SO2 in Upper Cloud Layer of Venus and Origin of UV-Absorption. Adv Sp Res 1981;1(9):13–16.

Zasova L V, Moroz VI and Linkin VM. Venera-15, 16 and VEGA Mission Results as Sources for Improvements of the Venus Reference Atmosphere. Adv Sp Res 1996;17(11):171–180.

Zhang X, Liang M-C, Montmessin F, et al. Photolysis of Sulphuric Acid as the Source of Sulphur Oxides in the Mesosphere of Venus. Nat Geosci 2010;3(12):834–837.

Zhang X, Liang MC, Mills FP, et al. Sulfur Chemistry in the Middle Atmosphere of Venus. Icarus 2012;217(2):714–739.

Zolotov MY and Matsui T. Chemical Models for Volcanic Gases on Venus. In: Lunar and Planetary Science Conference 2002; p. 1433.

Zolotov MY, Mogul R, Limaye SS, et al. Venus Cloud Composition Suggested From The Pioneer Venus Large Probe Neutral Mass Spectrometer Data. LPI Contrib 2023;2806:2880.